\documentclass[paper,11pt]{article}
\pdfoutput=1

\usepackage{mathrsfs}
\usepackage{graphicx}
\usepackage{epstopdf}
\usepackage{color}
\usepackage[centertags]{amsmath}
\usepackage{amsfonts}
\usepackage{amssymb} 
\usepackage{amsthm}
\usepackage{slashed}
\usepackage{framed}
\usepackage{enumerate,enumitem}
\usepackage{amsbsy}
\usepackage{jheppub}
\usepackage{subfig}
\usepackage{hyperref}

\def\({\left(}
\def\){\right)}
\def\[{\left[}
\def\]{\right]}

\def\one{{\rm 1\kern -.9mm l}}                             %

\def\beq{\begin{equation}}
\def\eeq{\end{equation}}
\def\beqa{\begin{eqnarray}}
\def\eeqa{\end{eqnarray}}
\newcommand{\eqa}{\begin{eqnarray}}
\newcommand{\ena}{\end{eqnarray}}

\def\bQ{{\bf Q}}

\newcommand{\Log}{\text{ln}}

\newcommand\blank[1]{}

\newcommand{\bP}{ {\bf P } }

\newcommand{\balpha}{\alpha\kern -6.7pt\alpha}
\newcommand{\bbalpha}{\alpha\kern -4.95pt\alpha}

\newcommand\en{\end{equation}}
\newcommand\bea{\begin{eqnarray}}
\newcommand\eea{\end{eqnarray}}
\newcommand\nn{\nonumber}

\newcommand{\One}{{\hbox{{\rm 1{\hbox to 1.5pt{\hss\rm1}}}}}}
\renewcommand{\One}{{\mathbb 1}}
\renewcommand{\One}{{\rm 1\!\!1}}

\newcommand{\IIm}{\text{Im}}
\newcommand{\ba}{\begin{eqnarray}}
\newcommand{\ea}{\end{eqnarray}}

\newcommand{\be}{\begin{equation}}
\newcommand{\ee}{\end{equation}}

\renewcommand{\log}{\ln}

\renewcommand{\ell}{{\mathcal L}}

\newcommand{\mA}{\mathcal{A}}
\newcommand{\sigtw}{\mathcal{P} }

\setlist[itemize]{leftmargin=*}
\def\XXint#1#2#3{{\setbox0=\hbox{$#1{#2#3}{\int}$ }
\vcenter{\hbox{$#2#3$ }}\kern-.6\wd0}}

\title{Exploring the spectrum of planar $AdS_4/CFT_3$ at finite coupling}

\author[1]{Diego Bombardelli}
\author[1,2]{,\;Andrea Cavagli\`a}
\author[1]{,\;Riccardo Conti}
\author[1]{and Roberto Tateo}

\affiliation[1]{Dipartimento\ di Fisica and INFN, Universit\`a di Torino, Via P.\ Giuria 1, 10125 Torino, Italy.}
\affiliation[2]{Mathematics Department, King's College London, The Strand, London WC2R 2LS, UK}

\emailAdd{dbombard@to.infn.it}
\emailAdd{andrea.cavaglia@kcl.ac.uk}
\emailAdd{conti@to.infn.it}
\emailAdd{tateo@to.infn.it}

\abstract{
The Quantum Spectral Curve (QSC) equations for  planar   $\mathcal{N}=6$ super-conformal Chern-Simons (SCS)   
are solved numerically at finite values of the coupling constant for states in the $\mathfrak{sl}(2|1)$ sector.

New weak coupling results for conformal dimensions of operators outside the $\mathfrak{sl}(2)$-like sector are  obtained by adapting a recently proposed  algorithm for the QSC perturbative  solution.
Besides being interesting in their own right, these perturbative results are necessary initial inputs for the numerical algorithm to converge on the correct solution.

The non-perturbative numerical outcomes nicely interpolate between the weak coupling and the known semiclassical expansions, and novel strong coupling exact results are deduced from the numerics. Finally, the existence of contour crossing singularities in the TBA equations for the operator $\textbf{20}$ is ruled out by our analysis.

The results of this paper are an important test of the QSC formalism for this model, open the way to new quantitative studies and provide further evidence in favour of the conjectured weak/strong coupling duality between  $\mathcal{N}=6$ SCS and type IIA superstring theory on $ AdS_4 \times CP^3$. Attached to the arXiv submission, a {\tt Mathematica} implementation of the numerical method and ancillary files containing the numerical results are provided.

}

\begin{document}
\maketitle

\section{Introduction}
The  first concrete realisation of the  $AdS/CFT$ duality was proposed in \cite{Maldacena:1997re,Gubser:1998bc,Witten:1998qj} and concerned the weak/strong coupling equivalence between the $\mathcal{N}=4$ super Yang-Mills (SYM) theory and the type IIB superstring theory on $AdS_5\times S^5$. 
For the current purposes, a crucial step was the  discovery of a link with integrability, at both weak and strong coupling \cite{MZ,Bena:2003wd}, in the planar ('t Hooft) limit of the  duality. 

Triggered by the works \cite{MZ,Bena:2003wd}, the spectrum of the theory was studied by adopting very powerful  integrable model techniques, such as the Bethe Ansatz (BA) \cite{MZ, Beisert:2005fw, Beisert:2006ez}, the Thermodynamic Bethe Ansatz (TBA) \cite{BFT,GKKV,AF} and closely related sets of functional relations \cite{Extended,Balog:2011nm,Wronskian,FiNLIE} which allowed to recast the spectral problem into a finite dimensional non-linear Riemann-Hilbert problem,  the Quantum Spectral Curve (QSC) \cite{QSC, Gromov:2014caa} (see \cite{Gromov:2017blm} and \cite{Kazakov:2018ugh} for recent reviews).
 
The QSC is probably  the ultimate simplification of the spectral  problem, as it allows to compute numerically the spectrum at finite coupling with high precision \cite{Gromov:2015wca,Hegedus:2016eop} and to inspect analytically interesting regimes such as the BFKL limit \cite{QCDPomeron,Gromov:2015vua} or the weak coupling expansions \cite{Marboe:2014gma,Marboe:2017dmb}.
 It has also been generalised\footnote{Besides, outside the AdS/CFT context this method was applied to simplify the formulation of thermodynamics for the Hubbard model \cite{Cavaglia:2015nta}. } to the $\gamma$ and $\eta$ deformations \cite{Thook, Klabbers:2017vtw}, to the so-called fishnet theory \cite{Gromov:2017cja}, to the quark-antiquark potential \cite{QSCCusp,QSCPotential} and very recently also to the calculation of correlators of three cusps in a special limit of $\mathcal{N}=4$ SYM \cite{Cavaglia:2018lxi}. 

Despite the considerable progress made in this research field, there are still many interesting open problems and possible generalisations,  see e.g. \cite{Gromov:2017blm}. 
 A practical problem  is the fact that, while the QSC potentially allows to study the  anomalous dimension of arbitrary operators, it is  rather  difficult to find  starting points ensuring the convergence of the iterative algorithm on a given chosen operator. Usually weak coupling data can be used efficiently as an initial seed for the numerics. An initial step towards  covering of the full spectrum of SYM was taken in \cite{Marboe:2017dmb}, solving the QSC at one loop for a wide set of states. 
 However, while at weak coupling there exists an efficient procedure to solve the QSC even beyond 10 loops \cite{ Marboe:2014gma,Gromov:2015vua}, at strong coupling a systematic perturbative approach is still missing (see however \cite{Hegedus:2016eop} for progress in this direction).

Planar $\mathcal{N}=4$ SYM is not the only interesting  $AdS/CFT$-related
integrable theory. Further examples, that for different reasons  can be considered
equally important, exist and concern supersymmetric conformal gauge theories in lower
space-time dimensions.
As a matter of fact, these models are intrinsically more complicated, are not maximally supersymmetric, and are
currently  much less understood compared to $\mathcal{N}=4$ SYM. 

The $AdS_4 /CFT_3$ case --- the main subject of the current paper --- was introduced by
Aharony, Bergman, Jafferis and Maldacena (ABJM) in  \cite{Aharony:2008ug} and is potentially
very important since it involves, on the $AdS$ part of the correspondence, a 4D quantum theory of gravity.

In the integrable planar limit the gauge side of the duality corresponds to the $\mathcal{N}=6$ superconformal Chern-Simons theory, while the gravity side is described by the type IIA superstring theory on $AdS_4 \times  CP^3$ \cite{Minahan:2008hf, Gaiotto:2008cg, Stefanski:2008ik, Arutyunov:2008if, Gromov:2008bz} (see also the review \cite{Klose:2010ki}). 

In contrast to  $\mathcal{N}=4$ SYM, in the ABJM theory integrability leaves unfixed the interpolating function $h(\lambda)$ \cite{Gaiotto:2008cg, Grignani:2008is}, which parametrises the dispersion relation of elementary spin chain/worldsheet excitations and enters as an effective coupling constant in the integrability-based approach. An important conjecture for the exact form of this function was made in \cite{Gromov:2014eha} by a comparison with the structure of localization results. The conjecture was extended in \cite{Cavaglia:2016ide} to the more general ABJ theory \cite{Aharony:2008gk}. The proposal of \cite{Cavaglia:2016ide} suggests that the only difference between ABJM and ABJ  corresponds to the replacement of $h(\lambda)$ with an explicitly defined $h^{\text{ABJ}}(\lambda_1, \lambda_2)$, where $\lambda_1, \lambda_2$ are the two (apparently) independent couplings of the ABJ theory. Therefore the analysis performed in this paper is also potentially relevant to the more general ABJ model.

Anomalous dimensions of single trace operators with asymptotically large quantum numbers are described, at all loops, by the Asymptotic Bethe Ansatz (ABA) equations, conjectured in \cite{Gromov:2008qe} and derived from the exact worldsheet S-matrix of \cite{Ahn:2008aa}. From the S-matrix, it was possible to obtain the leading order finite-size corrections (see for example \cite{Bombardelli:2008qd,Lukowski:2008eq,Ahn:2010eg,Beccaria:2010kd,Abbott:2011tp}). The exact result, including all finite-size corrections for short operators, is formally described by the infinite set of TBA equations and corresponding functional relations, proposed in \cite{Bombardelli:2009xz,Gromov:2009at,Cavaglia:2013hva}. The latter equations were solved numerically in \cite{LevkovichMaslyuk:2011ty} and the anomalous dimension of the operator \textbf{20} \cite{Minahan:2008hf} computed up to $h=1$. The results of \cite{LevkovichMaslyuk:2011ty} were so far the only  finite coupling data  for the spectrum of this model. 

The QSC equations of the ABJM model \cite{Cavaglia:2014exa, Bombardelli:2017vhk} have been already used to compute the so-called slope function in a near-BPS finite coupling regime \cite{Gromov:2014eha} and to develop an efficient algorithm for the weak coupling expansion in the $\mathfrak{sl}(2)$-like sector \cite{Anselmetti:2015mda}.

The main purpose of this work is to compute the finite coupling spectrum of a set of short operators by solving numerically the QSC equations. The  TBA results of \cite{LevkovichMaslyuk:2011ty} for the operator {\bf 20} are extended far beyond $h=1$, until the dual string description starts to emerge clearly, and a finite coupling analysis of other states in the $\mathfrak{sl}(2|1)$ sector is performed. This work can be considered as the first step toward a more systematic study of the ABJM finite coupling spectrum \cite{ComplexToappear}. 
As an attachment to the arXiv submission, we provide a simple Mathematica implementation of the numerical method for the subsector of parity-even operators; more general versions of the code are available upon request.

The rest of the paper is organized as follows. Section \ref{sec:review} contains a review of the basic QSC equations derived in \cite{Cavaglia:2014exa, Bombardelli:2017vhk}. The structure of the numerical algorithm is schematically described in section \ref{sec:algorithm} where the differences with respect to the $\mathcal{N}=4$ SYM case are underlined. In section \ref{sec:sl2} the numerical results obtained for two of the simplest and most studied operators in the $\mathfrak{sl}(2)$-like sector are reported and compared with existing finite coupling results \cite{LevkovichMaslyuk:2011ty} and strong coupling predictions \cite{Beccaria:2012qd, Gromov:2014eha}. Furthermore, the analytic structure of the function $\textbf{Y}_{1,0}$ associated to the operator \textbf{20} is carefully investigated, extending the computation performed with TBA techniques in \cite{LevkovichMaslyuk:2011ty} to larger values of the coupling constant and showing the absence of critical values of $h$ for this state. See for example \cite{Arutyunov:2009ax,Frolov:2010wt} for a discussion concerning the possible emergence of critical values of the coupling constant in $\mathcal{N}=4$ SYM.

In section \ref{sec:sl21}, the numerical analysis is extended to a couple of operators that do not belong to the $\mathfrak{sl}(2)$-like sector. One of the novel feature here is the emergence of a non-trivial $h$-dependent phase in the QSC equations, {\it i.e.} $\sigtw(h)$, which is explicitly computed both numerically at finite coupling and perturbatively at weak coupling. The results of this paper are an important test of the self-consistency of the QSC formulation of \cite{Bombardelli:2017vhk} even for these more complicated operators. 

The paper ends with a series of concluding remarks and four appendices which contain technical details about the symmetries of the QSC equations,  the reconstruction of the TBA solution from the Q functions,  analytic weak coupling expansions 
and the numerical results for $\Delta(h)$.

\section{Review of useful equations}
\label{sec:review}

In this section we review the basics of the QSC formulation presented in \cite{Cavaglia:2014exa, Bombardelli:2017vhk}. It involves a large number of Q functions, depending on the spectral parameter which we denote as $u$. Among them a primary role is played by the Q functions denoted as $\bP$ and $\bQ$, which can be viewed as a quantum version of the  quasi-momenta parametrising classical  string solutions in $AdS_4 \times CP^3 $. $\bP$ and $\bQ$ functions roughly correspond to $CP^3$ and $AdS_4$ degrees of freedom, respectively.   

\begin{figure}
\begin{minipage}{0.45 \linewidth}
\centering
\includegraphics[scale=0.3]{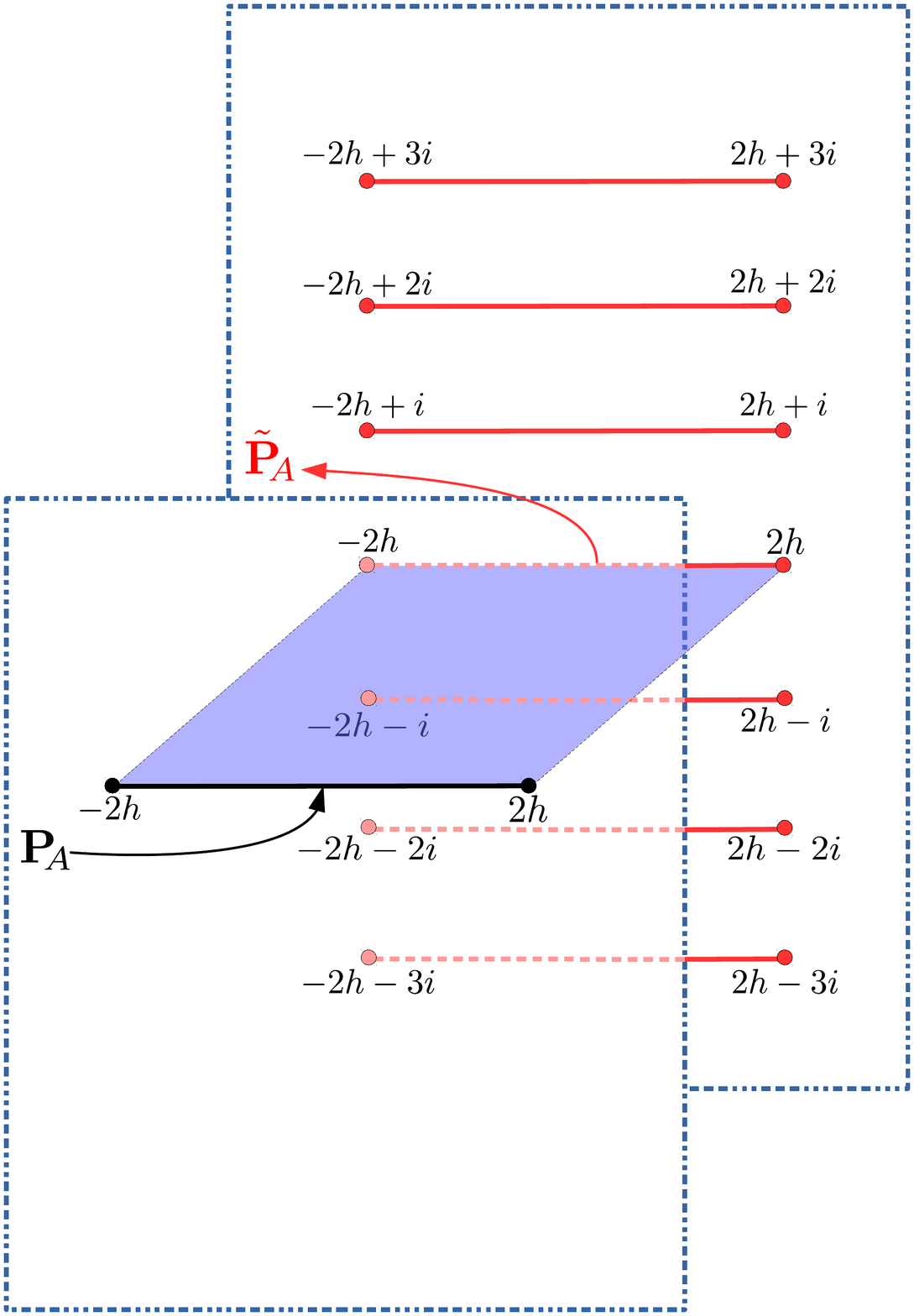}
\end{minipage}
\hspace{0.2cm}
\begin{minipage}{0.45 \linewidth}
\includegraphics[scale=0.3]{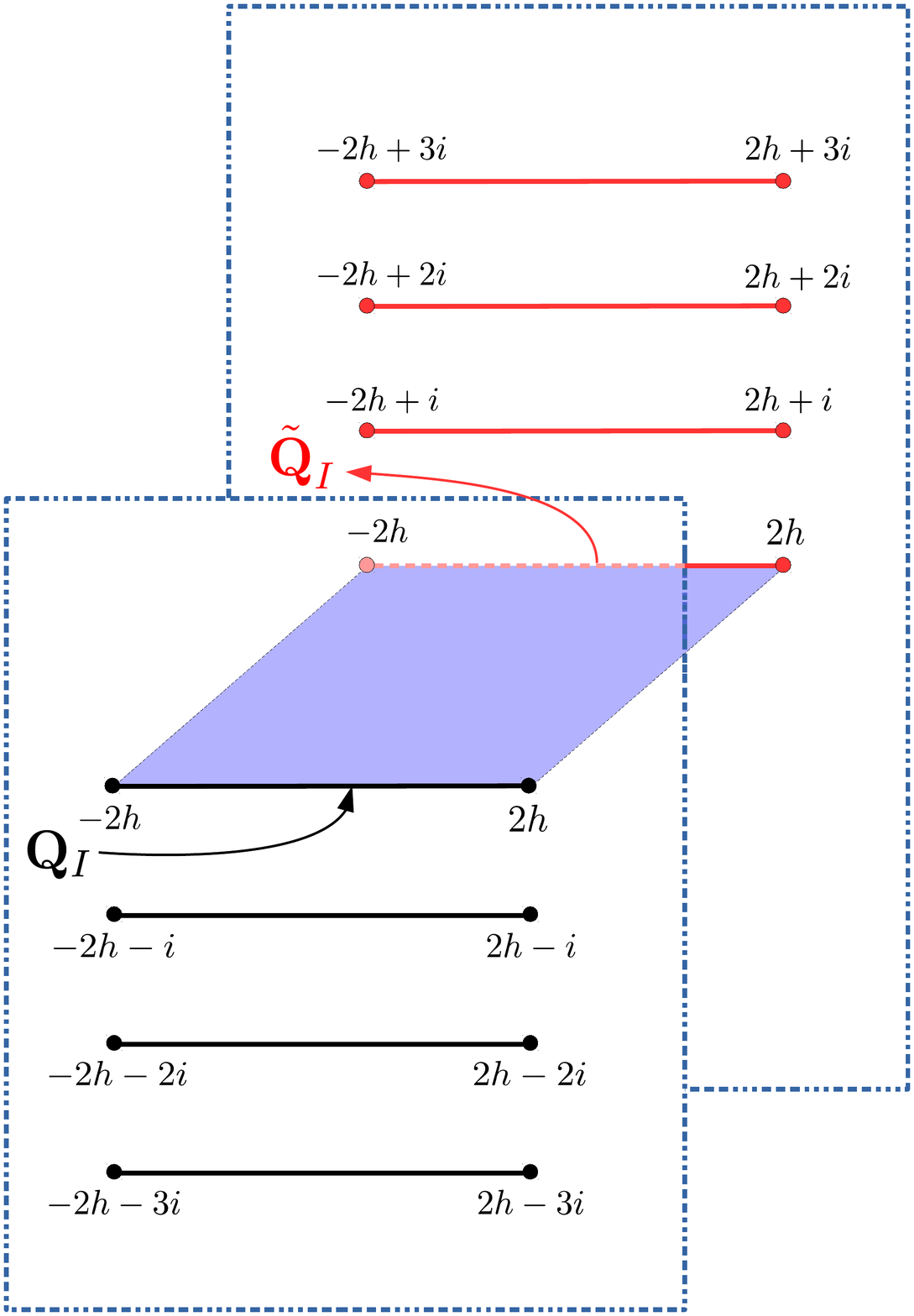}
\end{minipage}
\caption{Cut structure of the $\bP$ and $\bQ$ functions in the physical Riemann section. 
}
\label{fig:cutPQ}
\end{figure}

The $\bP$ functions enter a self-consistent formulation of the spectral problem, the $\bP\nu$-system, which is a closed set of discontinuity relations --- a  non-linear Riemann-Hilbert problem --- involving a finite number of unknown functions
\beqa
\{\bP_A(u)\}_{A=1}^6 \;,\; \{\nu_a(u), \nu^a(u) \}_{a=1}^{4} \;,
\eeqa
defined on a Riemann surface with an infinite number of sheets. On the reference Riemann sheet, the $\bP$'s  have a single, square-root type branch cut running from $-2h$ to $2h$. 
The $\nu$'s instead are required to fulfil the following  quasi-periodicity condition:
\beq
\label{eq:perioanti}
{ \widetilde \nu }_a(u)  = e^{i\sigtw} \, \nu_a(u+i) \;,\; { \widetilde \nu }^a(u)  =  e^{-i\sigtw} \, \nu^a(u+i) \;,
\eeq
where $\widetilde{f}(u)$ denotes the analytic continuation of $f(u)$ to the next sheet through the cut $u\in(-2h, 2h)$. 
In (\ref{eq:perioanti}), $\sigtw(h)$ is a state-dependent phase that may be, in general, a non-trivial function of the coupling constant $h$, as will be discussed in more detail in section \ref{sec:calP}. 
Setting
\beqa\label{eq:Pabdefd}
{ \bf P }_{ab} =   \left( \begin{array}{cccc} 0 & - \bP_1 & - \bP_2 & - \bP_5 \\  \bP_1 & 0 & - \bP_6 & - \bP_3 \\ \bP_2 & \bP_6 & 0 & - \bP_4 \\ \bP_5 & \bP_3 & \bP_4 & 0 \end{array} \right) \;,\;
{ \bf P }^{ab} =   \left( \begin{array}{cccc} 0 & - \bP^1 & - \bP^2 & - \bP^5 \\  \bP^1 & 0 & - \bP^6 & - \bP^3 \\ \bP^2 & \bP^6 & 0 & - \bP^4 \\ \bP^5 & \bP^3 & \bP^4 & 0 \end{array} \right) \;,
\eeqa
with $(\bP^1,\bP^2, \bP^3,\bP^4,\bP^5,\bP^6) = (-\bP_4,\bP_3,\bP_2,-\bP_1,-\bP_6,-\bP_5)$
and the constraints 
\beqa
{ \bf P }_{ac} { \bf P }^{cb} = \delta_{ab} \longleftrightarrow \bP_5 \bP_6 - \bP_2 \bP_3 + \bP_1 \bP_4 = 1 \;,\; \nu^a \, \nu_a = 0 \;, 
\label{eq:constraint} 
\eeqa
the $\bP\nu$-system is written as
\beqa
{ \widetilde {\bP} }_{ab} - { \bf P }_{ab} &=& \nu_a { \tilde \nu }_b - \nu_b { \tilde \nu }_a \;,\quad { \widetilde {\bP} }^{ab} - { \bf P }^{ab} = -\nu^a { \tilde \nu }^b + \nu^b { \tilde \nu }^a \;,
\label{eq:Pnu1}\\
{\tilde \nu }_a &=& -{ \bf P }_{ab} \; \nu^b \;,\ \ \ \ \ \ \ \ \ \ \ \ \ \ \ \ \ {\tilde \nu }^a = -{ \bf P }^{ab} \; \nu_b \;.
\label{eq:Pnu2}
\eeqa
 Notice that, as a consequence of these equations, the $\bP$  functions exhibit an infinite sequence of evenly-spaced short  cuts  from the second sheet onward with branch points located at $u=\pm 2h+i\mathbb{Z}$ (see figure \ref{fig:cutPQ}). 
The $\nu$ functions possess evenly-spaced short cuts with branch points at $u=\pm 2h+i\mathbb{Z}$ on all the Riemann sheets. The $\bP$'s and the $\nu$'s are required to be bounded and free of singularities, other than the branch points at $u = \pm 2 h + i \mathbb{Z}$, on every sheet of the Riemann surface. 

 In addition, equations (\ref{eq:constraint})--(\ref{eq:Pnu2}) 
 need to be supplemented with the following large-$u$ asymptotics of the $\bP$ functions:
\beq \label{eq:Pasy}
\bP_A(u) \sim \mathcal{A}_A \, u^{-M_A} \;,\; \bP^A(u) \sim \mathcal{A}^A \, u^{-M_A} \;,
\eeq
where
\beqa
\mathcal{A}_B \, \mathcal{A}^B = 2 \, \frac{\prod_{I=1}^5 \left(M_B - \hat{M}_I \right) }{ \prod_{C \neq B }^6 (M_B-M_C ) } \;,\; (B=1, \dots, 6)\;,
\label{eq:AAasy}
\eeqa
with no contraction over the index $B$. In (\ref{eq:Pasy}) and (\ref{eq:AAasy}), the charges  $M$ and $\hat{M}$ corresponding to a given state can be identified as (see \cite{Bombardelli:2017vhk} for more details)  
\beqa
\label{eq:MAdef}
M_A &=& \(L- K_3 + K_2 , \, L + K_1 -K_2 +1 , \, -M_2 , \,-M_1, \, -K_4 + K_{\bar4}, \,-M_5\) \;,\\
\hat{M}_I &=&  \left(\gamma + K_4 + K_{\bar{4}} + L - K_3 \,,\,\gamma +  L + K_1 +1\, ,\, -\hat{M}_2\, ,\, -\hat{M}_1 \, , \, 0\right)\;,
\label{eq:MIdef}
\eeqa
where $L$ is the spin chain length, $\gamma$ is the anomalous dimension and $K_i$ the excitation numbers in the ABA description of the states \cite{Gromov:2008qe,Ahn:2008aa}, in $\mathfrak{sl}(2)$ grading.
In particular, the following ordering holds: 
\beq\label{eq:asyorder}
|M_5| < M_1 < M_2  < \hat{M}_2 < \hat{M}_1 \; , 
\eeq 
which implements unitarity of the representation of the superconformal algebra. 

An important consequence of the analytic properties of the $\bP$ functions, which is fundamental for the numerical algorithm, is that they admit a convergent series representation:
\beq
\bP_{A}(u) = \frac{1}{\left(h\,x(u)\right)^{M_A}}\sum_{n=0}^{\infty} \frac{c_{A, n} }{ x^n(u) } \;,\; (A=1,\dots,6) \; , \label{eq:seriesP}
\eeq
where $x(u)$ is the Zhukovsky variable defined as
\beq
\label{eq:Zhukovsky}
x(u) = \frac{u + \sqrt{u- 2 h}\sqrt{u+2 h} }{2 h} \;.
\eeq
In principle, the set of equations (\ref{eq:constraint})--(\ref{eq:Pnu2}) already contains all the information necessary to compute the planar $AdS_4/CFT_3$ spectrum at fully non-perturbative level. However the currently available algorithms for the numerical solution at finite coupling of the $AdS_5/CFT_4$ spectrum \cite{Gromov:2015wca,Hegedus:2016eop} are based on other subsets of the QSC equations, which involve both $\bP$ and $\bQ$ functions.

For this purpose, we define 16 functions $Q_{a|i}(u) \; (a,i=1,\dots,4)$ \cite{Bombardelli:2017vhk}, as solutions of the $4$-th order finite difference equation\footnote{Throughout all the paper the shorthand notation $f^{[n]}(u)=f\left( u+ n\,i/2 \right)$ and $f^{\pm}(u)=f^{[\pm 1]}(u)$ will be adopted for shifts in the rapidity variable $u$. The shifts are assumed to be performed without crossing any of the cuts at $u\in(-2 h, 2h) + i \mathbb{Z}$. }
\beq
\label{eq:defQai}
Q_{a|i}^{[+2]} = \textbf{P}_{ab}^{+} \left( \textbf{P}^{bc} \right)^{-} Q_{c|i}^{[-2]} \hspace{1mm}.
\eeq 
The $Q_{a|i}$'s are required to be analytic in the whole upper half plane (they turn out to have an infinite ladder of short branch cuts in the lower half plane starting from $\IIm(u)=-1/2$), to have power-like asymptotics at large $u$ and can be normalised as
\beq
Q_{a|i}^+ \, \bP^{ab} \, Q_{b|j}^- = \kappa_{ij} ,
\eeq
where $\kappa_{ij}$ is an anti-symmetric matrix, independent of $u$, whose only nonzero entries are $\kappa_{14} = \kappa_{32} = -\kappa_{23} = - \kappa_{41} = 1$. They can be used to construct the $\bQ$  and $\tau$ functions --- the  $AdS_4$ counterpart of the $\bP$ and $\nu$ functions --- as
\beq
\label{eq:Qijdef}
\bQ_{ij} = - Q_{a|i}^- \bP^{ab} Q_{b|j}^- 
\;,\; \tau_i = \nu^a \, Q_{a|i}^- \;.
\eeq
The corresponding $\bQ\tau$-system is
\beqa\label{eq:QtilTautil}
&& \widetilde{\bQ}_{ij} - \bQ_{ij} =  \widetilde{\tau}_i \, \tau_j- \widetilde{\tau}_j \, \tau_i \;,\;
\widetilde{\tau}_i = - \bQ_{ij} \, \tau^j \; ,
\eeqa
with $\tau^i \equiv e^{- i \cal{ P } } \,  \kappa^{ij} \, \tau_j^{++} $, and 
$\kappa^{ij}$ is the inverse of $\kappa_{ij}$.  The cut structure of the $\bQ$ functions is represented in figure \ref{fig:cutPQ}, while the $\tau$'s  inherit 
from the $\nu$'s  the infinite set of   evenly-spaced short cuts at $u=\pm 2h+i\mathbb{Z}$ and have the $2i$-periodicity property $\tau_i^{[+4]}= \tau_i$.
 Below, we will use a more convenient vector notation for the $\bQ$ functions: \beq
\label{eq:Qijcomp0}
\bQ_I =-\{ \bQ_{12} , \bQ_{13} ,  \bQ_{24} , \bQ_{34} , \frac{1}{2} \, (\bQ_{14} + \bQ_{23}) \} \;\;\;,\;\;\; \bQ_{\circ} = 2 \, ( \bQ_{23} - \bQ_{14} ) .
\eeq
 To summarise the large-$u$ asymptotics of the main Q functions it is convenient to introduce the combination of charges
{ \small
\beqa
\label{eq:NaNi}
\mathcal{N}_a &=& \( \frac{1}{2} (-M_1-M_2 -M_5) \,,\, \frac{1}{2} (-M_1+M_2+M_5) \,,\, \frac{1}{2}(M_1-M_2 + M_5) \,,\, \frac{1}{2} ( M_1+M_2 - M_5 ) \)\;, \nn\\
\hat{\mathcal{N}}_i &=& \( \frac{1}{2} (\hat M_1 + \hat M_2 ) \,,\, \frac{1}{2} (\hat M_1-\hat M_2 ) \,,\, \frac{1}{2}(\hat M_2 - \hat M_1 ) \,,\, \frac{1}{2} (-\hat M_1-\hat M_2) \) \; , 
\eeqa }
which allow us to write
\begin{eqnarray}
\bP_{ab}(u) \sim u^{ \mathcal{N}_a + \mathcal{N}_b} \;&,& \bQ_{ij}(u) \sim u^{ \hat{\mathcal{N}}_i + \hat{\mathcal{N}}_j}
\;,\;\;\; Q_{a|i}(u) \sim u^{\mathcal{N}_a + \hat{\mathcal{N}}_i } \; .  \label{eq:Qaiasym}
\end{eqnarray}

Finally, upon a specific choice of basis for the solutions of the system (\ref{eq:defQai}), the $\bQ$ functions and their analytic continuations fulfil a further set of constraining equations, the so-called {\it gluing conditions},\footnote{See \cite{Gromov:2015vua} for a first derivation in the context of $AdS_5/CFT_4$.} which were derived in \cite{Bombardelli:2017vhk} for the ABJM model in the case of half-integer spin and real values of $h$.
These equations are the main ingredient of the numerical algorithm.
 For the validity of the gluing conditions, we choose a  solution of (\ref{eq:defQai}) with a particular large-$u$ asymptotic expansion of the ``pure" form \cite{Gromov:2015vua, Bombardelli:2017vhk}:
\begin{equation}
\label{eq:pure}
Q_{a|i}(u) \simeq u^{\mathcal{N}_a+\hat{\mathcal{N}}_i} \sum_{n = 0}^{\infty} \frac{B_{a|i,n}}{u^n} \; ,
\end{equation}
where, for the purposes of this paper, all coefficients  $B_{(a|i), n}$ are real. 
Then the gluing conditions are (see \cite{Bombardelli:2017vhk}):
\beqa
\label{eq:gluingconditions1}
\hspace{-0.7cm}\widetilde{\bQ}_{1} &=& -\frac{e^{ i \pi  \hat M_1}}{\cos( \pi  \hat M_1 ) } \, \overline{\bQ}_{1} + \delta_1 \, \overline{\bQ}_{3} \;,\;
 \widetilde{\bQ}_{3} = -\frac{e^{-i \pi \hat M_1}}{\cos( \pi \hat M_1 ) } \, \overline{\bQ}_{3} +  \delta_2 \, \overline{\bQ}_{1} \;, \\
\hspace{-0.7cm}\widetilde{\bQ}_{2} &=& - \frac{e^{i \pi \hat M_1}}{\cos( \pi \hat M_1 ) } \, \overline{\bQ}_{2} +   \delta_1 \, \overline{\bQ}_{4} \;,\;
 \widetilde{\bQ}_{4} = -\frac{e^{-i \pi \hat M_1}}{\cos( \pi \hat M_1 ) } \, \overline{\bQ}_{4} +\delta_2 \,  \overline{\bQ}_{2} \;, \label{eq:gluingconditions4}\\
\hspace{-0.7cm}\widetilde{\bQ}_{\circ} &=& \overline{\bQ}_{\circ} \;,\; \widetilde{\bQ}_{5} =  - \overline{\bQ}_{5} \;,
\label{eq:gluingconditions5}
\eeqa
where the coefficients $\delta_1$, $\delta_2$ are constrained by 
\beq
\label{eq:delta1delta2}
\delta_1 \, \delta_2 = \tan^2( \pi \hat{M}_1 ) \;.
\eeq

\section{The algorithm}
\label{sec:algorithm}

This section contains the description of the algorithm implemented for the numerical computation of the $AdS_4/CFT_3$ spectrum, using the equations reviewed in section \ref{sec:review}.

\subsection{General setup}
\label{sec:General Setup}

\begin{figure}
\begin{minipage}{0.45\textwidth}
\centering
\vspace{-1.7cm}
\includegraphics[scale=0.4]{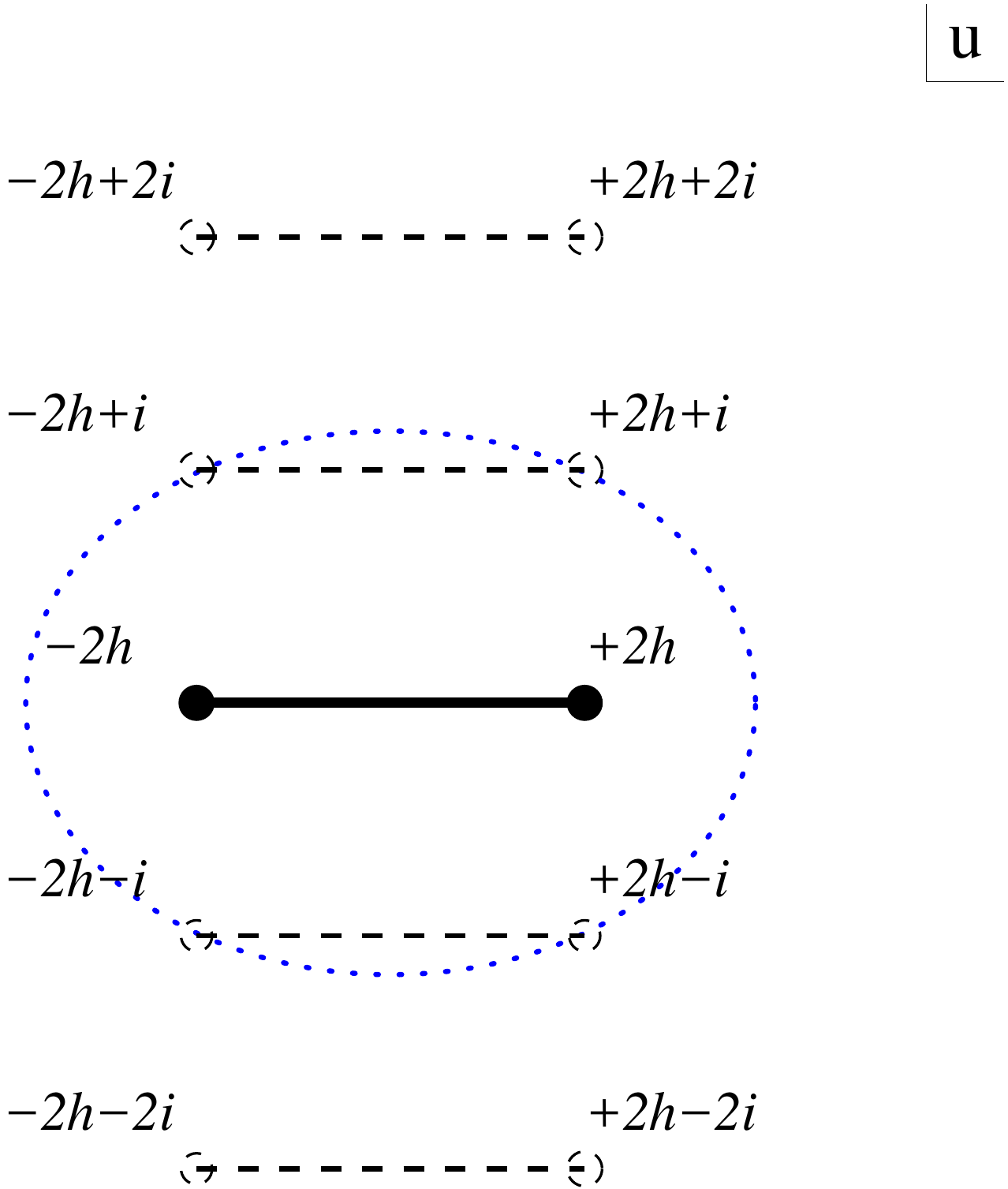}
\vspace{0.25cm}
\caption{Analytic structure of the $\textbf{P}$ functions in the $u$-plane. 
The dotted ellipse delimits the convergence region of the power expansions of $\textbf{P}_A$ in the second sheet of the $u$-plane.}
\label{fig:cutP}
\end{minipage}
\hspace{0.5cm}
\begin{minipage}{0.45\textwidth}
\includegraphics[scale=0.28]{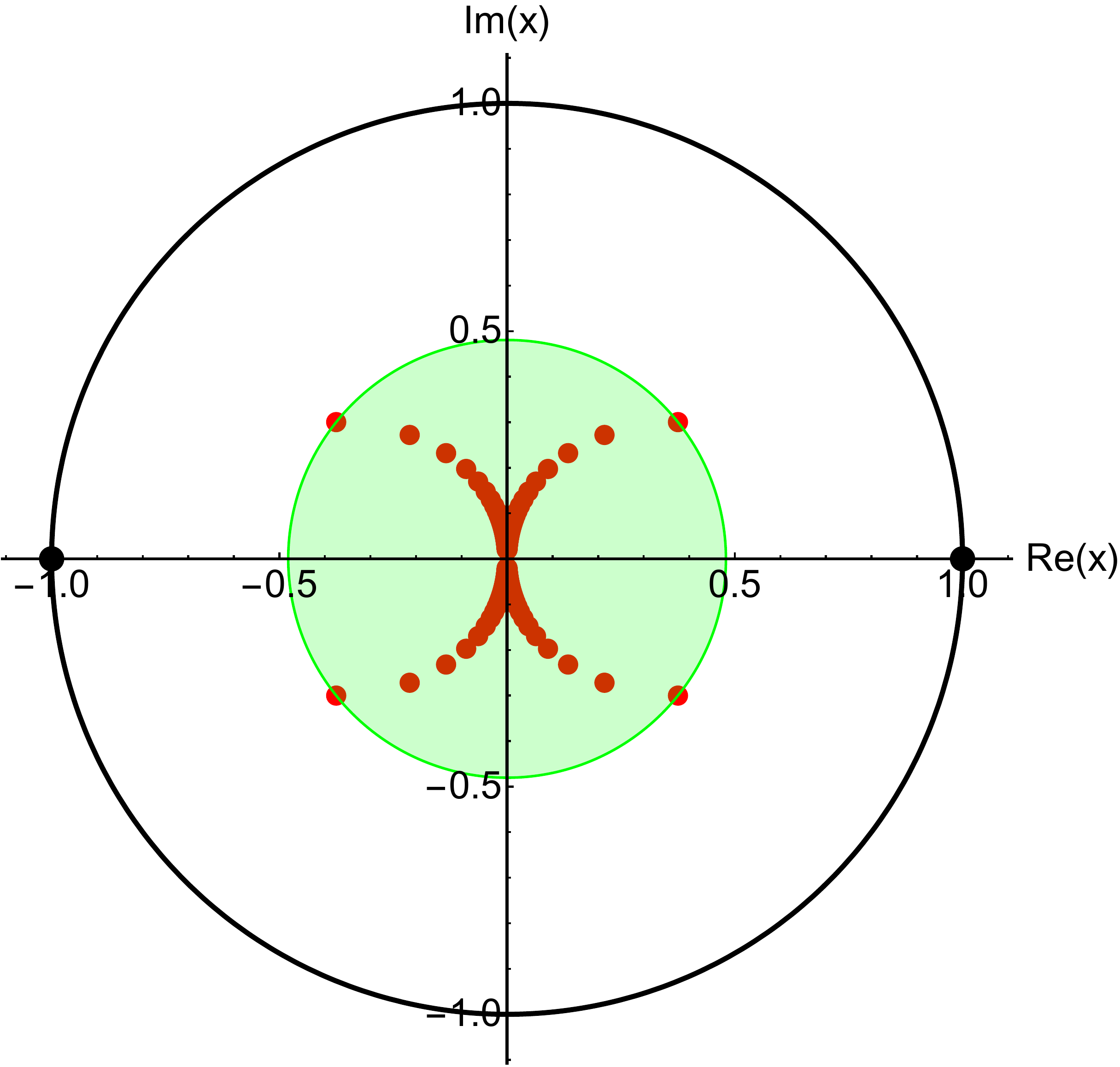}
\caption{Analytic structure of the $\textbf{P}$ functions in the $x$-plane for $h=1$. The unit circle separates the first sheet ($|x(u)|>1$) from the second sheet ($|x(u)|<1$). The inner green disc represents the convergence region of the power expansions of $\textbf{P}_A$, the red dots correspond to the positions of the branch points on the second sheet.}
\label{fig:convergence}
\end{minipage}
\end{figure} 

Following \cite{Marboe:2014gma}, we expand the $\textbf{P}$ functions in  powers of the Zhukovsky variable $x(u)$ around $x=\infty$ with coefficients $\left\lbrace c_{A,n} \right\rbrace_{n\geq 0}$ as in (\ref{eq:seriesP}). 

In analogy with the $\mathcal{N}=4$ SYM case \cite{Gromov:2015wca}, it is  simple to deduce from the analytic properties of the $\bP$ functions that (\ref{eq:seriesP}) converges everywhere on the first sheet of the $u$-plane, which corresponds to $|x(u)| > 1$ in figure \ref{fig:convergence}, and also in an elliptic region around the cut $u\in(-2h,2h)$ on the second sheet (see figure \ref{fig:cutP}). The convergence region  is indeed bounded by the position of the nearest branch points, which lie at $u=\pm2h\pm i$ on the second sheet. On the other hand, an analogous Laurent expansion in $x(u)$ for the $\textbf{Q}$ functions would not even converge on all points of the cut $u\in(-2 h, 2h)$. This is the reason why one starts with a series representation for the $\textbf{P}$'s and not for the $\textbf{Q}$'s. 

As will be discussed in the following sections and in \cite{ComplexToappear}, the convergence of (\ref{eq:seriesP}) is a particularly delicate issue at strong coupling and when analytic continuation to complex values of $h$ is numerically implemented. Since in the current context we will mainly adopt the $x$-plane perspective, it is useful to introduce here some general concepts. Formula (\ref{eq:Zhukovsky}) maps the complex plane of $u$ into the complex plane of $x$ according to the following rules
\begin{align*}
u-\textbf{plane} \hspace{2,5cm}&\hspace{2,5cm} x-\textbf{plane}\\
\text{linear cut from}\,-2h\;\text{to}\,+2h \hspace{1,0cm}&\hspace{2,5cm} \text{unit circle}\\
u=\infty\;\text{(first sheet)} \hspace{1,8cm}&\hspace{2,8cm} x=\infty\\
u=\infty\;\text{(second sheet)} \hspace{1,5cm}&\hspace{2,9cm} x=0
\end{align*}
and the whole second sheet in the $u$-plane is mapped into the interior of the unit circle. Thus the series (\ref{eq:seriesP}) converges everywhere in the $x$-plane except for the inner green disk in figure \ref{fig:convergence} containing all the branch points belonging to the second sheet. Since the gluing conditions are evaluated on the cut between $-2h$ and $2h$ (unit circle in figure \ref{fig:convergence}), it is crucial that the unit circle falls completely inside the convergence region of the $\textbf{P}$ functions. As $h$ is increased along the real axis, the branch points approach $x=\pm 1$ and correspondingly the numerical algorithm, which is based on a truncation of the series (\ref{eq:seriesP})
\begin{equation}
\label{eqn:truncated power expansion of P}
\textbf{P}_A (u) = \frac{1}{\left(hx(u)\right)^{M_A}}\sum_{n = 0}^{N_A} \frac{c_{A,n}}{x^n(u)} \;,\; (A = 1,\dots,6) \;,
\end{equation}
becomes less and less efficient, since the cutoff $N_A$ should be accordingly increased with $h$.

The final objective is to determine --- with very high numerical accuracy --- the set of coefficients $\left\lbrace c_{A,n} \right\rbrace_{n=0}^{N_A}$ solving the QSC equations together with the gluing conditions. Since the conformal dimension $\Delta$ is related to $c_{A,0}$ via (\ref{eq:Pasy})--(\ref{eq:MIdef}), in the following we will work with the equivalent set of unknowns
\beq
\vec{X} = \left\lbrace X_a \right\rbrace_{a=1}^{N_A+1} = \left( \Delta\,,\,\left\lbrace c_{A,n} \right\rbrace_{n = 1}^{N_A}\right) \;.
\eeq
Schematically, the algorithm consists of two main blocks.
In the first block $\textbf{Q}_I(u)\equiv\textbf{Q}_I(u;\vec{X})$ and $\widetilde{\textbf{Q}}_I(u)\equiv\widetilde{\textbf{Q}}_I(u;\vec{X})$ are formally evaluated on the cut $u\in(-2h,+2h)$ in terms of the parameters $\vec{X}$. This enters as a subroutine in the main part of the program, in which $\vec{X}$ is fixed by imposing the gluing conditions on the cut; for this purpose, the iterative Levenberg-Marquardt procedure is used.

\paragraph{Part 1: computing $\textbf{Q}_I$ and $\widetilde{\textbf{Q}}_I$.}
\label{sec:Qfunctions}

The strategy to compute $\textbf{Q}_I(u)$ and $\widetilde{\textbf{Q}}_I(u)$ on the cut $u\in(-2h,+2h)$ is the following. Starting from the ansatz~\eqref{eqn:truncated power expansion of P} for the $\textbf{P}$'s, first we compute $Q_{a|i}^+(u)$ on the cut by solving (\ref{eq:defQai}). Below, the method to solve such a finite difference equation is explained in detail.  
Then, using formula (\ref{eq:Qijdef}) and its analytic continuation
\beq
\textbf{Q}_{ij} = \left( Q^{a|k} \right)^{+} \kappa_{ki}\, \textbf{P}_{ab} \left( Q^{b|l} \right)^{+} \kappa_{lj} \;,\;
\widetilde{\textbf{Q}}_{ij} = \left( Q^{a|k} \right)^{+} \kappa_{ki}\, \widetilde{\textbf{P}}_{ab} \left( Q^{b|l} \right)^{+} \kappa_{lj} \;,
\eeq
we find $\textbf{Q}_{ij}(u)$ and $\widetilde{\textbf{Q}}_{ij}(u)$. 
The computation of the $\textbf{Q}$ functions is therefore reduced to the solution of equation (\ref{eq:defQai}) for $Q_{a|i}^+(u)$ on the segment $u\in(-2h,+2h)$; this is done in a two step calculation. First we find an approximate solution to (\ref{eq:defQai}) at some $u$ with large integer imaginary part $\text{Im}(u)=N_s$. For this purpose, we truncate the large-$u$ asymptotic series representation (\ref{eq:pure}): 
\begin{equation}
Q_{a|i}(u) \simeq u^{\mathcal{N}_a+\hat{\mathcal{N}}_i} \sum_{n = 0}^{N_{a|i}} \frac{B_{a|i,n}}{u^n} \;,
\label{eqn:large u Qai}
\end{equation}
which allows to reduce the finite difference equation (\ref{eq:defQai}) to a much simpler linear system\footnote{Let us stress that such linear system is safely solvable only if the components of $Q_{a|i}$ have definite parity in $u$: 
otherwise it is necessary to adopt the strategy explained in appendix \ref{app:resonance}.} for the unknowns $\left\lbrace B_{a|i,n} \right\rbrace_{n = 1}^{N_{a|i}}$, where the leading order coefficients $B_{a|i,0}$ are known up to a gauge choice.
Then, iterating $N_s$ times equation (\ref{eq:defQai}), the large-$u$ solution can be shifted down to $u\in(-2h,+2h)$:
\begin{equation}
Q_{a|i}\Bigl(u+\frac{i}{2}\Bigr) = \bigl[ \textbf{P}(u+i)\textbf{P}^{-1}(u+2i) \textbf{P}(u+3i)\dots \textbf{P}^{-1}(u+i N_s) \bigr]_{a}^{\,b} Q_{b|j}\Bigl(u + i N_s + \frac{i}{2}\Bigr) \;.
\label{eq:Qaishift}
\end{equation}

\paragraph{Part 2: fixing the coefficients $\vec{X}$.}
\label{sec:Newtonmethod}

Here the strategy is to fix $\vec{X}$ by imposing the gluing conditions on the cut $u\in(-2h,+2h)$. A suitable functional $\mathcal{F}(\vec{X})$ is built out of the gluing conditions, in such a way that it vanishes when the latter are satisfied. Then  $\vec{X}$ is obtained looking for a root of  $\mathcal{F}(\vec{X})$. 

 We begin by rewriting (\ref{eq:gluingconditions1})--(\ref{eq:gluingconditions5})
  in a discretised form
\beqa
& f_1(u_k;\vec{X}) =\widetilde{\textbf{Q}}_1(u_k)+\alpha \, \overline{\bQ}_{1}(u_k) - \delta_1 \, \overline{\bQ}_{3}(u_k) \;, \label{eq:gluingimpl1} &\\
&f_2(u_k;\vec{X}) =\widetilde{\textbf{Q}}_2(u_k)+\alpha \, \overline{\bQ}_{2}(u_k) - \delta_1 \, \overline{\bQ}_{4}(u_k) \;, \label{eq:gluingimpl2} &\\
&f_3(u_k;\vec{X})=\widetilde{\textbf{Q}}_3(u_k)+\beta \, \overline{\bQ}_{3}(u_k) - \delta_2 \, \overline{\bQ}_{1}(u_k) \;, \label{eq:gluingimpl3} &\\
&f_4(u_k;\vec{X})=\widetilde{\textbf{Q}}_4(u_k)+\beta \, \overline{\bQ}_{4}(u_k) -\delta_2 \,  \overline{\bQ}_{2}(u_k) \;, \label{eq:gluingimpl4} &\\
&f_5(u_k;\vec{X})=\widetilde{\bQ}_{5}(u_k) + \overline{\bQ}_{5}(u_k) \;, \label{eq:gluingimpl5}  &\\
&f_6(u_k;\vec{X})=\widetilde{\bQ}_{\circ}(u_k) - \overline{\bQ}_{\circ}(u_k) \;, \label{eq:gluingimpl6}&
\eeqa
together with their analytic continuation $\left\lbrace \tilde{f}_i(u_k;\vec{X}) \right\rbrace_{i=1}^6$ obtained by replacing $\widetilde\bQ_I\rightarrow \bQ_I$ and $\overline\bQ_I\rightarrow\widetilde{\overline\bQ}_I$. In (\ref{eq:gluingimpl1})--(\ref{eq:gluingimpl6}) the coefficients are defined as
\beq
\alpha = \frac{e^{ i \pi  \hat M_1}}{\cos( \pi  \hat M_1 ) } = e^{ 2i \pi  \hat M_1} \beta \;,\;\;\;\;\; \delta_1 = \tan^2( \pi \hat M_1 )/\delta_2 , 
\eeq
while $\left\lbrace u_k \right\rbrace_{k=1}^{N_p}$ give a discretisation of the  interval $(-2 h, 2h)$. For numerical convergence the optimal choice of discretisation points corresponds to the zeros of the Chebyshev polynomials of the first kind adapted to the interval, {\it i.e.} $u_k = 2 h \cos(\frac{2 k-1}{2 N_p} ) \;,\; (k=1, \dots, N_p)$.

Notice that the gluing conditions (\ref{eq:gluingconditions1})--(\ref{eq:gluingconditions5}) are defined up to the parameter $\delta_2$,  which cannot be fixed a priori. This ambiguity is lifted by treating it as a genuine additional unknown:
\beq
\vec{X} = \left\lbrace X_a \right\rbrace_{a=1}^{N_A+2} = \left\lbrace  \delta_2 \,,\Delta\,,\,\left\lbrace c_{A,n} \right\rbrace_{n = 1}^{N_A}\right\rbrace \;,
\eeq
leaving equations (\ref{eq:gluingimpl1})--(\ref{eq:gluingimpl6}) formally unchanged\footnote{ In \cite{Gromov:2015wca} a slightly different method is used to remove a similar ambiguity in the $\mathcal{N}=4$ SYM gluing conditions, namely the unfixed parameters are evaluated by constructing normalisation independent combinations of Q  functions. 
While this is perfectly equivalent, we found the method described here to be numerically more stable in some challenging regimes, such as at strong coupling and in the proximity of the branch point at $h \sim i/4$ \cite{ComplexToappear}.}.

The next step is to arrange (\ref{eq:gluingimpl1})--(\ref{eq:gluingimpl6}) into a $(12\,N_p)$-dimensional vector
\beq
\vec{f}(\vec{X}) = \left\lbrace f_I(\vec{X}) \right\rbrace_{I=1}^{12\,N_p} = \left\lbrace f_i (u_k;\vec{X}) \right\rbrace \cup \left\lbrace \tilde{f}_i (u_k;\vec{X}) \right\rbrace \;,\; (i=1,\dots,6 \; ;\; k=1,\dots,N_p) \;,
\eeq
where the generic element is labelled by the multi-index $I=(i,k)$. It is then natural to define the functional $\mathcal{F}(\vec{X})$ as the squared norm of the vector $\vec{f}(\vec{X})$, {\it i.e.}
\beq
\mathcal{F}(\vec{X}) = |\vec{f}(\vec{X})|^2 = \sum_{I=1}^{12\,N_p} f_I(\vec{X})\bar{f}_I(\vec{X}) \;.
\label{eqn:functional}
\eeq
$\mathcal{F}(\vec{X})$ is a real and positive defined quantity which vanishes when the gluing conditions are fulfilled.

The Levenberg-Marquardt method appears to be the right choice for a minimisation problem of this kind, as observed in \cite{Gromov:2015wca, Hegedus:2016eop}. To implement this iterative procedure efficiently, an initial guess --- close enough to the solution --- for the parameters $\vec{X}^{(0)}$ is needed. For small values of $h$ up to $h\simeq 0.30$, analytic data from the weak coupling expansions\footnote{For the symmetric operators discussed in section \ref{sec:L1S1} and \ref{sec:L1S2} we used the analytic results of \cite{Anselmetti:2015mda} up to 8 loops, whereas for the operators studied in section \ref{sec:nonsym} and \ref{sec:L02} weak coupling data are computed by generalising the same method, based on the $\textbf{P}\nu$-system, to non-symmetric sectors (see appendix \ref{app:Pweak} for the explicit results).} provide a good starting point for $\Delta$ and $\left\lbrace c_{A,n} \right\rbrace_{n = 1}^{N_A}$. Whereas to move outside the weak coupling regime in practice it is necessary to change the coupling in small steps, using an extrapolation to obtain the initial configuration for a given value $h$.

Similarly to \cite{Gromov:2015wca}, for most of the operators discussed in the following, we find that for the convergence of the algorithm it is sufficient to impose the validity of a subset of the gluing conditions: in particular, $f_i(u_k;\vec{X})$ with $i=3,5,6$ are found to be sufficient in most cases. Instead, an additional gluing condition, {\it i.e.} $f_2(u_k,\vec{X})$ or equivalently $f_4(u_k,\vec{X})$, is necessary for the numerical convergence of the method in the non-symmetric case discussed in section \ref{sec:L02}. This is so far an experimental observation and it would be interesting to clarify why this is the case.

Concerning the convergence of the iterative algorithm, it turns out to be very important to reduce the space of the parameters to a submanifold where the solution of the QSC is  non-degenerate. For example, the quadratic constraint (\ref{eq:constraint}) was imposed at each iteration of  (\ref{eq:defQai}) by considering $\bP_4$ as a function of the other five $\bP$ functions throughout, therefore reducing the set of independent parameters to $\left\{ c_{A, n} \right\}_{n=1}^{N_A}$ with $A \neq 4$. Furthermore, the QSC  admits a continuous family of symmetries, which implies that infinitely many different sequences of coefficients $\left\{ c_{A, n} \right\}$ may be  used for the  description of  the same physical state. The gauge fixing of these extra symmetries is discussed in detail in appendix \ref{app:sym}.

Finally, before discussing the applications of the algorithm, let us focus briefly on the precision of the numerical results, which is mainly affected by the cut-off parameters:
\begin{itemize}
\item the truncation order in the power series~\eqref{eqn:truncated power expansion of P}, $N_A$;
\item the truncation order in the asymptotic series~\eqref{eqn:large u Qai}, $N_{a|i}$;
\item the imaginary part of the large $u$ approximation, $N_s$;
\item the number of sampling points, $N_p$.
\end{itemize}

Table \ref{tab:ruletime} displays specific values of the cut-off parameters and the computing times corresponding to some of the results listed in table \ref{tab:DeltaL1S2} of appendix \ref{app:numericalsl2}. The algorithm is implemented in a {\tt Mathematica} notebook using a processor with 16 cores at 2.10 GHz each and 32 GB of RAM.
The precision of the results was estimated by considering the number of stable digits, written in between brackets, as the truncation parameters are slightly increased. 

\begin{table}
\begin{equation*}
\begin{array}{|ccccccc|}
\hline
h & N_A & N_{a|i} & N_s & N_p & \#\;\text{of decimal digits}\; & \;\text{computing time}\\ 
\hline
0.10 & 50 & 38 & 50 & 52 & 30 & \sim 40\, \mbox{mins} \\ 
0.50 & 72 & 40 & 72 & 74 & 31 & \sim 1\, \mbox{hour}\\ 
1.00 & 76 & 52 & 76 & 78 & 28 & \sim 1.5\, \mbox{hours}\\ 
2.00 & 92 & 68 & 92 & 94 & 22 & \sim 3.5\, \mbox{hours}\\ 
3.00 & 110 & 78 & 110 & 112 & 21 & \sim 9\, \mbox{hours}\\ 
\hline
\end{array}
\end{equation*}
\caption{Set of parameters used to get some of the conformal dimensions of the $\mathfrak{sl}(2)$-like operator with $L=1$, $S=2$.}
\label{tab:ruletime}
\end{table}

\section{Spectrum at finite coupling: $\mathfrak{sl}$(2)-like operators}
\label{sec:sl2}

The Bethe Ansatz Equations (BAEs) describing the asymptotic spectrum of the ABJM $\mathfrak{sl}(2)$-like sector are \cite{Gromov:2008qe}
\beq
\label{eq:BAEsl2}
\left(\frac{x_{4,k}^+}{x_{4,k}^-}\right)^{L} =   \prod_{j=1}^{S}\frac{u_{4,k}-u_{4,j}-i}{u_{4,k}-u_{4,j}+i} \,\left( \frac{1-\frac{1}{x_{4,k}^+x_{4,j}^-} }{1-\frac{1}{x_{4,k}^-x_{4,j}^+}} \, \sigma_{}(u_{4,k},u_{4,j})\right)^2 \;,
\eeq
where $x_{4, j}=x(u_{4, j} )$ and $\sigma$ is the BES dressing factor \cite{Beisert:2006ez}.
The BAEs (\ref{eq:BAEsl2}) have to be supplemented by the zero momentum condition (ZMC)
\beq
\label{eq:zmcsl2}
\left(\prod_{k=1}^S\frac{x_{4,k}^+}{x_{4,k}^-}\right)^2=1\,.
\eeq
The states described by equations (\ref{eq:BAEsl2}) and (\ref{eq:zmcsl2}) correspond to single-trace operators of the form
\beq
\text{tr}\[D_+^S(Y^1Y_4^{\dagger})^L\] \;,
\eeq
and do not form a proper closed sector of the theory, but rather a collection of states within the wider $\mathfrak{sl}(2|1)$ sector\footnote{See for example \cite{Papathanasiou:2009zm} and \cite{Klose:2010ki} for a more detailed discussion of this point.}.
The ABA predictions for the conformal dimension of these operators is
\beq
\label{eq:DeltaABA}
\Delta_{\mathfrak{sl}(2)}^{ABA}=L+S+\sum_{k=1}^{S}\left(\sqrt{1+16\,h^2\sin^2\left(\frac{p_k}{2}\right)}-1\right)\,,\ \mbox{with}\ e^{ip_k}=\frac{x_{4,k}^+}{x_{4,k}^-}\,.
\eeq
The large-$u$ asymptotics of the $\bP$ functions, that are one of the main initial inputs of the algorithm, can be easily selected, for this sector, by setting $K_1=K_2=K_3=0$ and $K_4=K_{\bar{4}}=S_{}$ in (\ref{eq:MAdef}). 
The result is \cite{Cavaglia:2014exa}
\beq
\label{eq:asymsl2}
\bP_A\sim\left(\mathcal{A}_1u^{-L},\mathcal{A}_2u^{-L-1},\mathcal{A}_3u^{L+1},\mathcal{A}_4u^{L},\mathcal{A}_5u^0,\mathcal{A}_6u^0\right)\, ,
\eeq
 and besides the symmetry $\bP_6 = \bP_5$ can be imposed.  

\subsection{The {\emph L} = 1, {\emph S} = 1 operator}
\label{sec:L1S1}

 \begin{figure}
\centering
 \includegraphics[width=0.65\textwidth]{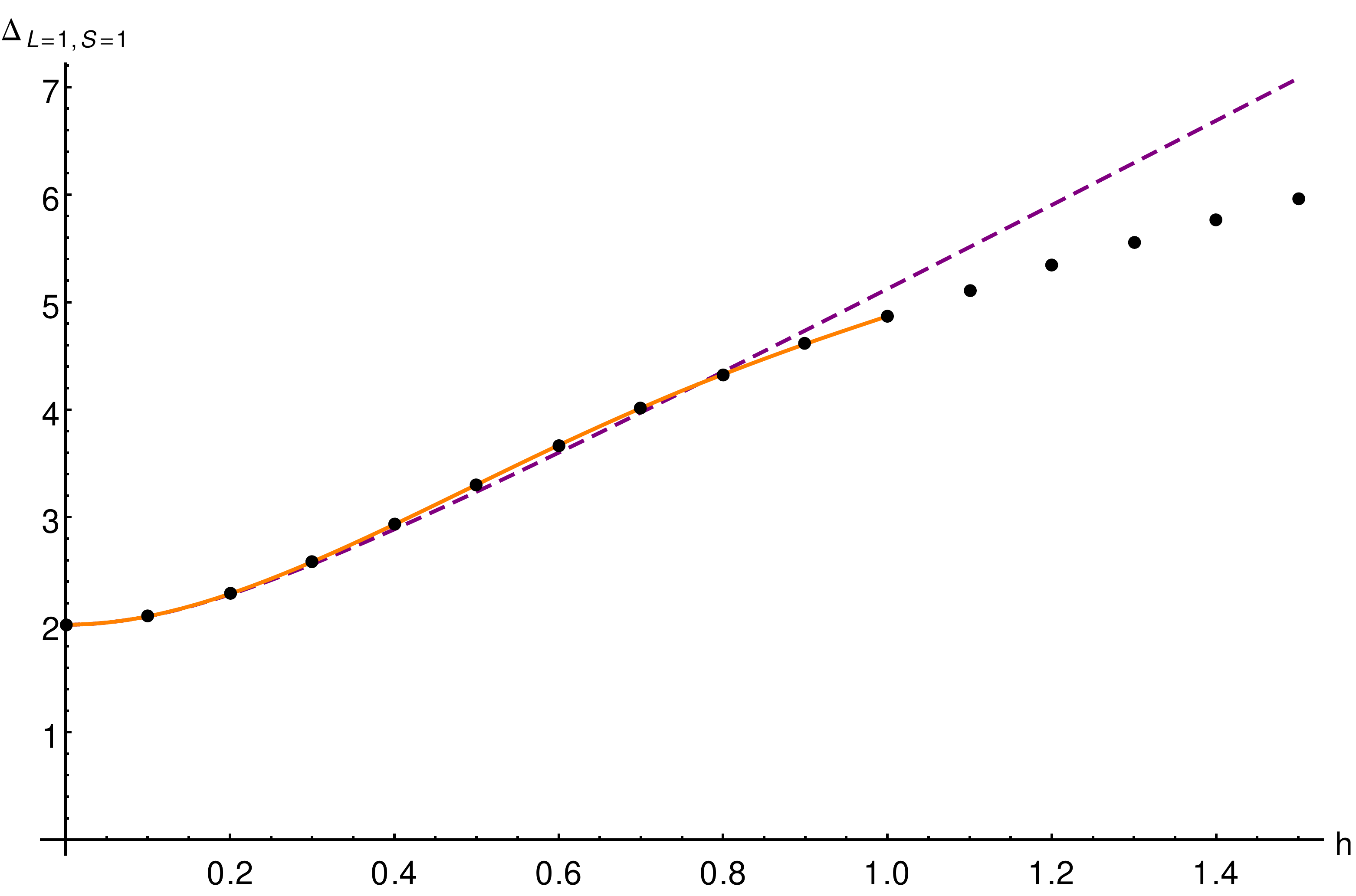}
\caption{Plot of $\Delta_{L=1,S=1}(h)$ for $h\in(0,1.5)$: the dots correspond to our numerical data, the dashed line represents the all-loop ABA expression (\ref{eq:deltaABAL1S1}) and the solid line interpolates the TBA results of \cite{LevkovichMaslyuk:2011ty}. For a plot in a wider range of $h$, see figure \ref{fig:L1S1}.}
\label{fig:L1S1zoom}
\end{figure}

The simplest non-protected operator belonging to the $\mathfrak{sl}(2)$-like sector is the operator {\bf 20} \cite{Minahan:2008hf, Gromov:2009tv, Beccaria:2009ny, Minahan:2009aq, Minahan:2009wg, Beccaria:2010kd, Anselmetti:2015mda} with length $L=1$ and spin $S=1$. The BAEs (\ref{eq:BAEsl2})--(\ref{eq:zmcsl2}) reduce to
\beq
\frac{x_{4,1}^+}{x_{4,1}^-}=-1\;,
\label{eq:abaL1S1}
\eeq
and the corresponding all-loop asymptotic conformal dimension is
\beq
\Delta_{L=1,S=1}^{ABA} = 1+\sqrt{1+16\,h^2} \;.
\label{eq:deltaABAL1S1}
\eeq
 The first non-perturbative numerical study was performed in \cite{LevkovichMaslyuk:2011ty} by solving numerically the TBA \cite{Bombardelli:2009xz, Gromov:2009at} up to $h=1$. The results obtained using the QSC-based algorithm described in section \ref{sec:algorithm} are reported in appendix \ref{app:numericalsl2}, table \ref{tab:DeltaL1S1}. 
 As shown in figure \ref{fig:L1S1zoom}, the TBA data are consistent with our results and give an important independent test of the correctness of the method. However, the numerical precision of the data obtained  in this paper with the QSC is much higher. 

Since the main motivation for the study of this model resides in the weak/strong coupling $AdS/CFT$ duality, it is particularly important to explore the large $h$ behaviour of the spectrum.  Strong coupling predictions  for operators in the $\mathfrak{sl}(2)$ sector are based on analytic continuation from the  classical folded spinning string solution  \cite{Beccaria:2012qd, Gromov:2014eha}, and is expected to be applicable only to operators with even $S$. Therefore, we found no independent  strong coupling predictions for the operator $\bf 20 $. However, the high precision results in table \ref{tab:DeltaL1S1} allow us to extract a numerical prediction for the first few strong coupling coefficients of $\Delta_{L=1,S=1}$. By analogy with the $S$ even case, we shall assume the following ansatz \cite{Gromov:2015wca} 
\beq
\label{eq:strongansatz}
\Delta = \sum_{n=0} \Delta^{(n)} g^{\frac{1-n}{2}}
\;,
\eeq
where $g=2\pi h+\log 2$. This is a natural parameter for the strong coupling expansion (\ref{eq:strongansatz}), since $g \sim \sqrt{( \lambda - 1/24 )/2}$ at large $\lambda$, where the shift $-1/24$ is expected from string theory considerations \cite{Bergman:2009zh,Gromov:2014eha}.

\begin{table}
\begin{center}
\def\arraystretch{1.3}
\begin{tabular}{|c||c|c|c|c|c|c|}
\hline
 $n$ & $\Delta^{(n)}_{\mathrm{fit}}$ & $\Delta^{(n)}_{\mathrm{guess}}$ & $|\Delta^{(n)}_{\mathrm{fit}}-\Delta^{(n)}_{\mathrm{guess}}|$\\
 \hline
 0 & 1.99999(3) & 2 & 7.2$\times 10^{-6}$ \\
 1 & -0.49999(7) & $-\frac{1}{2}$ & 2.9$\times 10^{-6}$ \\
 2 & 0.56250(0) & $\frac{9}{16}=0.5625$ & 1.1$\times 10^{-7}$ \\
 3 & -2.7837(2) & $-\frac{81}{1024}-\frac{9\zeta_3}{4}=-2.78372959\dots$ & 9.3$\times 10^{-6}$ \\
  \hline
\end{tabular}

\caption{Strong coupling coefficients for $L=1$, $S=1$. }
\label{tab:coeffL1S1}
\end{center}
\end{table}

\begin{figure}
\centering
 \includegraphics[width=0.65\textwidth]{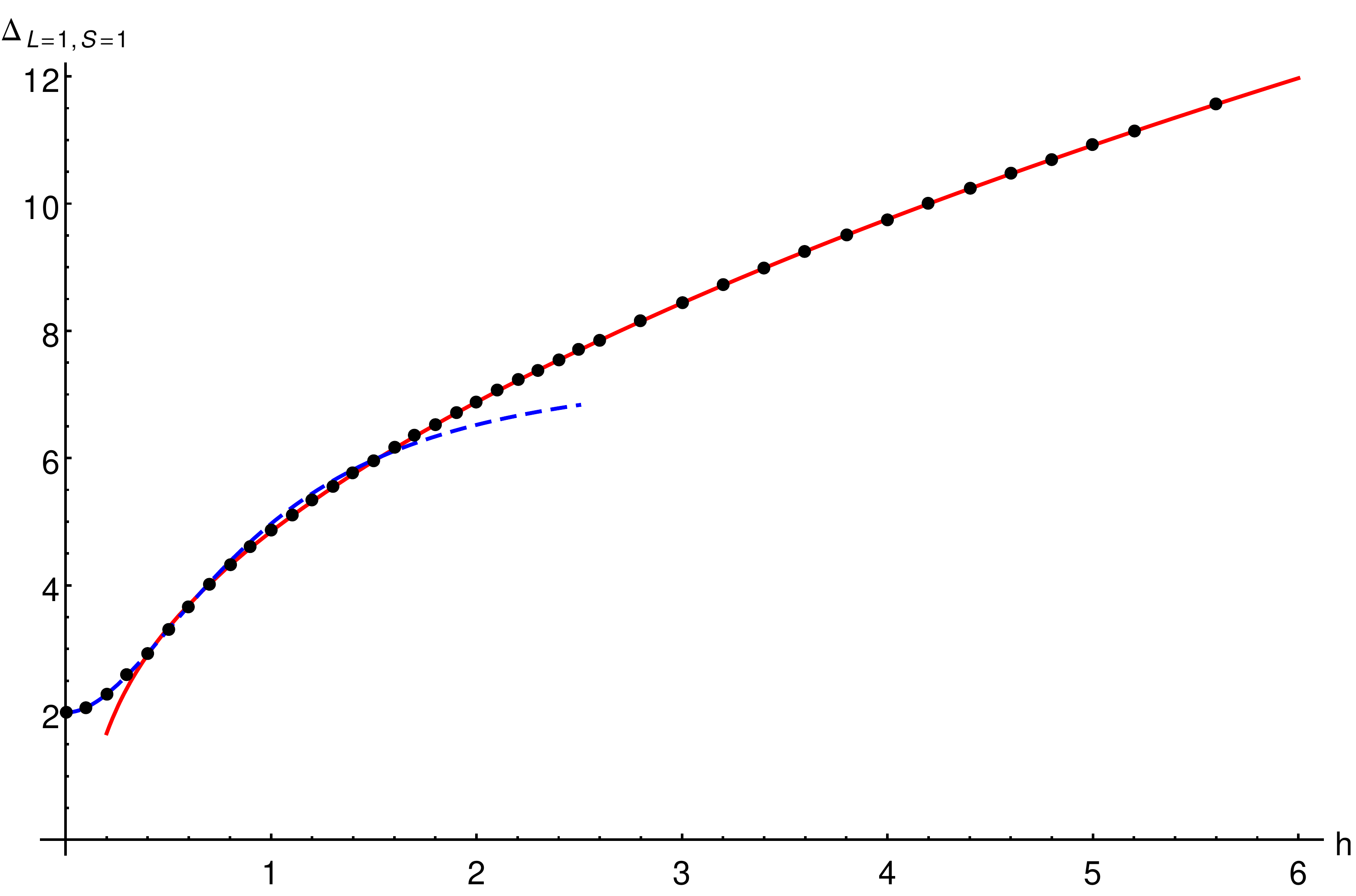}
\caption{Plot of $\Delta_{L=1,S=1}(h)$: the dots correspond to the numerical data reported in table \ref{tab:DeltaL1S1} of appendix \ref{app:numericalsl2}, while the solid line and the dashed line represent the strong coupling expansion (\ref{eq:strongprediction}) and the weak coupling Pad\'e approximant (\ref{eq:PadeL1S1}) respectively.}
  \label{fig:L1S1}
 \end{figure}

Table \ref{tab:coeffL1S1} contains the numerical strong coupling coefficients $\Delta_{\mathrm{fit}}^{(n)}$ obtained by fitting the results in table \ref{tab:DeltaL1S1} with the ansatz (\ref{eq:strongansatz}). Following the method of \cite{Hegedus:2016eop}, the coefficients $\Delta_{\mathrm{fit}}^{(n)}$ are obtained by increasing the truncation order in (\ref{eq:strongansatz}) while keeping the number of interpolating data fixed,
 until the result 
stabilises. The associated uncertainty corresponds to the last stable digit, written between brackets in table \ref{tab:coeffL1S1}. The resulting coefficients are in very good agreement with $\Delta_{\mathrm{guess}}^{(n)}$, corresponding to the following analytic expression
\beq
\label{eq:strongprediction}
\Delta_{L=1,S=1}= 2\,\sqrt{g} -\frac{1}{2} + \frac{9}{16\,\sqrt{g}}-\left(\frac{81}{1024}+\frac{9\zeta_3}{4}\right)\frac{1}{g^{3/2}}+\mathcal{O}\left(\frac{1}{g^{5/2}}\right) \;.
\eeq
To guess the fourth coefficient $\Delta_{\mathrm{guess}}^{(3)}$ we assumed a certain similarity with the known result for $S$ even \cite{Beccaria:2012qd, Gromov:2014eha}, see also section \ref{sec:L1S2}.
We stress that already the leading order coefficient in (\ref{eq:strongprediction}) deviates from the ones naively obtained interpolating the even-$S$ results, confirming that this operator belongs to a different trajectory. It would be interesting  to reproduce (\ref{eq:strongprediction}) by an analytic computation, identifying the appropriate family of classical solutions. 

It is important to  remark here that, while the strong coupling limit of the ABA formula (\ref{eq:deltaABAL1S1}) gives
\beq
\Delta_{L=1, S=1}^{ABA}=\frac{2}{\pi}\,g+\mathcal{O}(1)\,,
\eeq
the leading order of (\ref{eq:strongprediction}) matches instead the expectations of \cite{LevkovichMaslyuk:2011ty}, 
and the known $\lambda^{1/4}$ strong coupling behaviour of analogous anomalous dimensions in $\mathcal{N}=4$ SYM \cite{Gubser:1998bc}. 

In figure \ref{fig:L1S1}, a  nice overlap of the numerical data with both (\ref{eq:strongprediction}) and a diagonal $[6/6]$ Pad\'e approximant of the weak coupling expansion up to 12 loops \cite{Anselmetti:2015mda}, is observed. The Pad\'e prediction is:
\beq
\label{eq:PadeL1S1}
\Delta_{L=1, S=1}^{\mbox{\scriptsize Pad\'e}}=\frac{2 + 50.5387
   \,h^2 + 369.8384
  \,h^4 + 735.3660
   \,h^6}{1 + 21.2693
 \,h^2 + 113.0012
   \,h^4 + 97.8284
   \,h^6}\;.
\eeq 

\subsubsection{Study of the TBA function {\bf Y}\textsubscript{1,0}}
\label{sec:Y10}

\begin{figure}[t]
  \centering
  \includegraphics[scale=0.35]{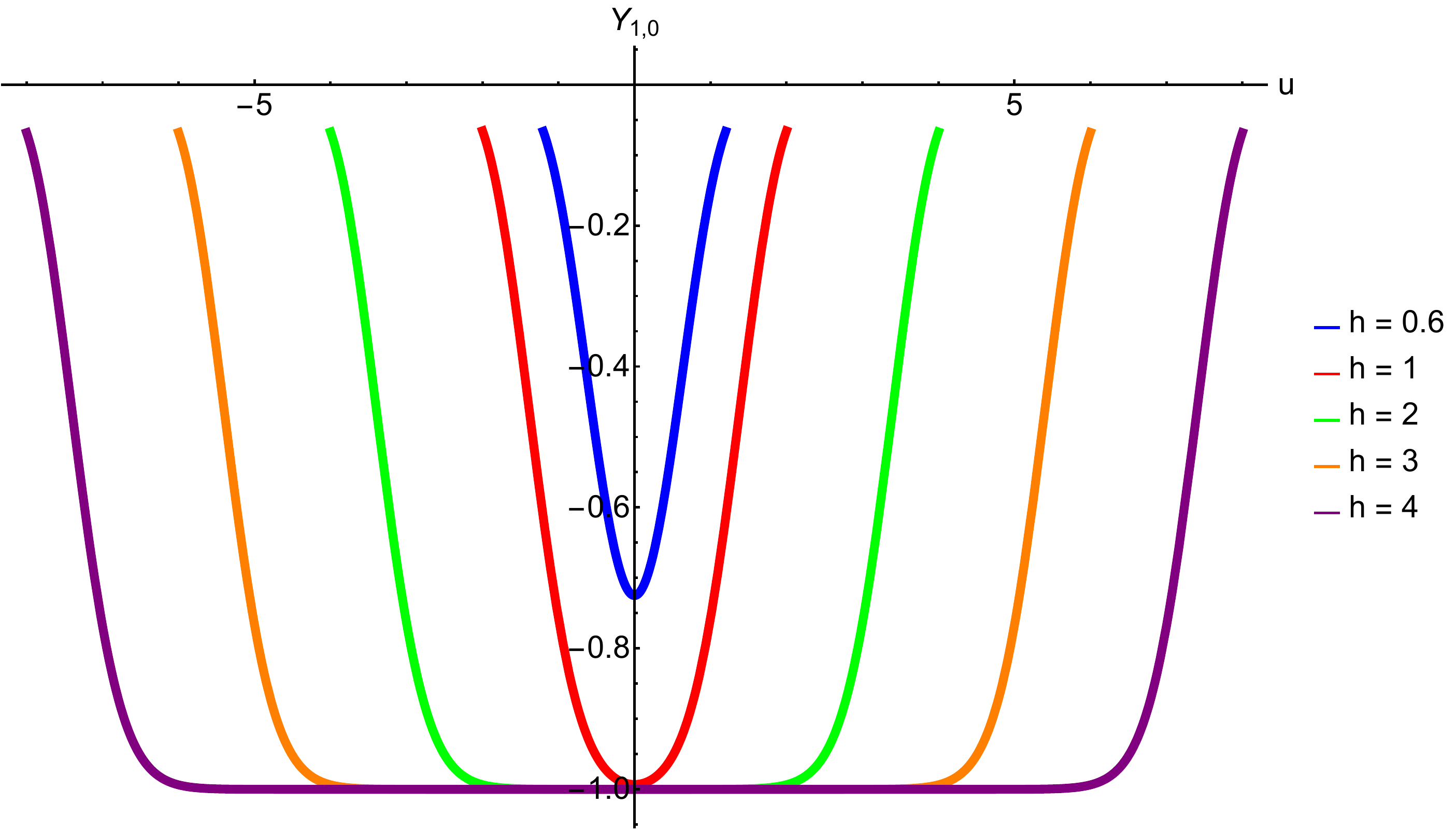}
  \caption{Plot of $\textbf{Y}_{1,0}(u)$ for the $L=1,S=1$ operator, with $u\in(-2h,+2h)$ and different values of $h$. It seems clear that $\textbf{Y}_{1,0}(u)$ tends to the constant value $-1$ as the coupling grows.}
  \label{fig:Y10RealAxis}
  \end{figure}

As noted  in \cite{LevkovichMaslyuk:2011ty}, the solution of the TBA becomes numerically unstable beyond $h=1$. In the TBA setup, the unknowns are the so-called $\textbf{Y}$ functions and the origin of the instability was  identified by the author of \cite{LevkovichMaslyuk:2011ty}  in the apparently divergent behaviour of $\log\left(1+\textbf{Y}_{1,0}(u)\right)$ around $h=1$.

The aim of this section is to investigate numerically the behaviour of the function $\textbf{Y}_{1,0}$ for the operator {\bf 20} at strong coupling using the QSC-based algorithm. In doing so we will answer an interesting question on the behaviour of this function raised in \cite{LevkovichMaslyuk:2011ty}.

As was noted in \cite{LevkovichMaslyuk:2011ty}, $\textbf{Y}_{1,0}(0)$ gets closer and closer to the value $-1$  as the coupling is increased toward $h\simeq 1$. In view of the structure of the TBA equations, which involve convolution integrals over $\log{(1+\textbf{Y}_{1,0}(u))}$ , it is quite interesting to determine if $\textbf{Y}_{1,0}$  actually crosses the value $-1$   as the coupling increases further. In fact, in the presence of singularities crossing the integration contour, the TBA should be  modified 
with the inclusion of extra residue terms \cite{Dorey:1996re} or, equivalently,  by implementing  a ``desingularisation'' procedure  that guarantees the correct  analytic continuation  of the physical state \cite{Bazhanov:1996aq,DoreyPerturbedCFT}. Speculation on the existence of such critical values in the context of the TBA for $\mathcal{N}=4$ SYM was presented in \cite{Arutyunov:2009ax,Frolov:2010wt}.

\begin{figure}[h]  
\centering
  \includegraphics[scale=0.35]{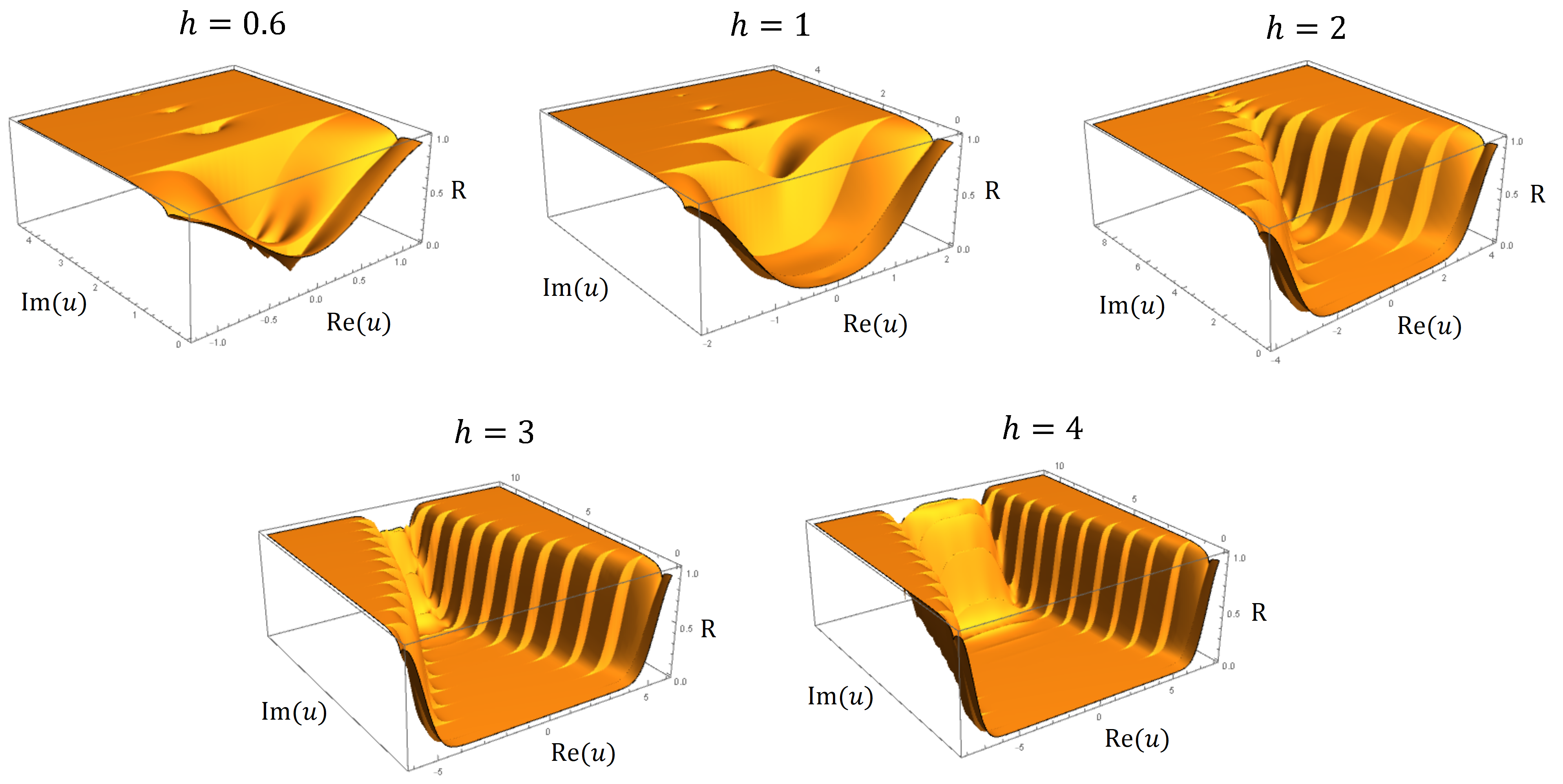}
  \caption{Plots of the ratio $R$, defined in (\ref{eq:R}), for different values of $h$. $R$ tends to zero in the whole complex plane as $h$ grows, thus hinting that ${\bf Y}_{1,0}$ tends to $-1$ at strong coupling, for the $L=1,S=1$ operator.}
  \label{fig:Y10}
\end{figure}

The function $\textbf{Y}_{1,0}(u)$ can be reconstructed efficiently starting from the numerical solution of the QSC, which in fact allows to compute all Y functions \cite{Bombardelli:2017vhk}. For technical details see appendix \ref{app:Y}. 
It is noteworthy that, while the approach of the value $-1$ causes a severe instability in the TBA equations, it is harmless from the point of view of the QSC, as it simply corresponds to the zero of a Q function approaching the cut. 
 
From the numerical outcomes displayed in figure \ref{fig:Y10RealAxis}, we see that $\textbf{Y}_{1,0}$ develops a wider and wider plateau as $h$ is increased while $(1+\textbf{Y}_{1,0})$ remains positive. These numerical results strongly suggest that there are no contour-crossing singularities of the kind $\textbf{Y}_{1,0}(u)=-1$  for any finite value of $h$.

Finally, it is also interesting to investigate the analytic structure of $\textbf{Y}_{1,0}(u)$ as $u$ is continued to the complex plane. Since $\textbf{Y}_{1,0}(u)$ is in general a complex-valued function for $u\in\mathbb{C}$, we studied the real ratio
\beq
R(u)=\frac{|1+\textbf{Y}_{1,0}^{-1}(u)|}{1+|1+\textbf{Y}_{1,0}^{-1}(u)|} \;,
\label{eq:R}
\eeq
in which the values $\textbf{Y}_{1,0}= (-1,0)$ are mapped into $R= (0,1)$. From figure \ref{fig:Y10} it clearly appears that $\textbf{Y}_{1,0}(u)$ tends to the constant value $-1$ in the whole complex plane, asymptotically as $h$ tends to infinity.

\subsection{The  {\emph L} = 1, {\emph S} = 2 operator}
\label{sec:L1S2}

Since an analytic strong coupling expansion is available for all the $\mathfrak{sl}(2)$ states with even $S$ \cite{Beccaria:2012qd, Gromov:2014eha}, it is 
interesting to analyse also the operator with length $L=1$ and spin $S=2$. In this case the BAEs (\ref{eq:BAEsl2})--(\ref{eq:zmcsl2}) reduce to $u_{4,2}=-u_{4,1}$ and
\beq
\frac{x_{4,1}^+}{x_{4,1}^-}=   -\frac{2u_{4,1}-i}{2u_{4,1}+i} \,\left( \frac{1+\frac{1}{(x_{4,1}^+)^2} }{1+\frac{1}{(x_{4,1}^-)^2}} \, \sigma(u_{4,1},-u_{4,1})\right)^2 \;.
\label{eq:abaL1S2}
\eeq 

The numerical results obtained for the conformal dimension are reported in table \ref{tab:DeltaL1S2} of appendix \ref{app:numericalsl2}, and plotted in figure \ref{fig:L1S2} together with a $[6/6]$ Pad\'e approximant of the weak coupling expansion up to $12$ loops \cite{Anselmetti:2015mda}:
\beq
\label{eq:PadeL1S2}
\Delta_{L=1,S=2}^{\mbox{\scriptsize Pad\'e}}=\frac{3+66.0075
 \,  h^2+382.5693
 \,  h^4+481.1827
 \,  h^6}{1+19.3358
\, h^2+88.2894
 \,  h^4+72.9268
 \,  h^6}\;.
\eeq 
In figure \ref{fig:L1S2}, we also  plotted the strong coupling predictions obtained from the ansatz (\ref{eq:strongansatz}) in \cite{Beccaria:2012qd, Gromov:2014eha}:
\beq
\label{eq:strongpredictionL1S2}
\Delta_{L=1,S=2}= 2\,\sqrt{g} -\frac{1}{2} + \frac{25}{16\,\sqrt{g}}+\left(\frac{271}{1024}-\frac{9\zeta_3}{4}\right)\frac{1}{g^{3/2}}+\mathcal{O}\left(\frac{1}{g^{5/2}}\right) \;.
\eeq
Table \ref{tab:coeffL1S2} contains the numerical predictions for the strong coupling coefficients --- obtained by fitting the results in table \ref{tab:DeltaL1S2} --- which are in very good agreement with (\ref{eq:strongpredictionL1S2}). 
This result can be considered as a further strong evidence of the gauge/string duality involving the ABJM theory.

\begin{figure}
\centering
 \includegraphics[width=0.6\textwidth]{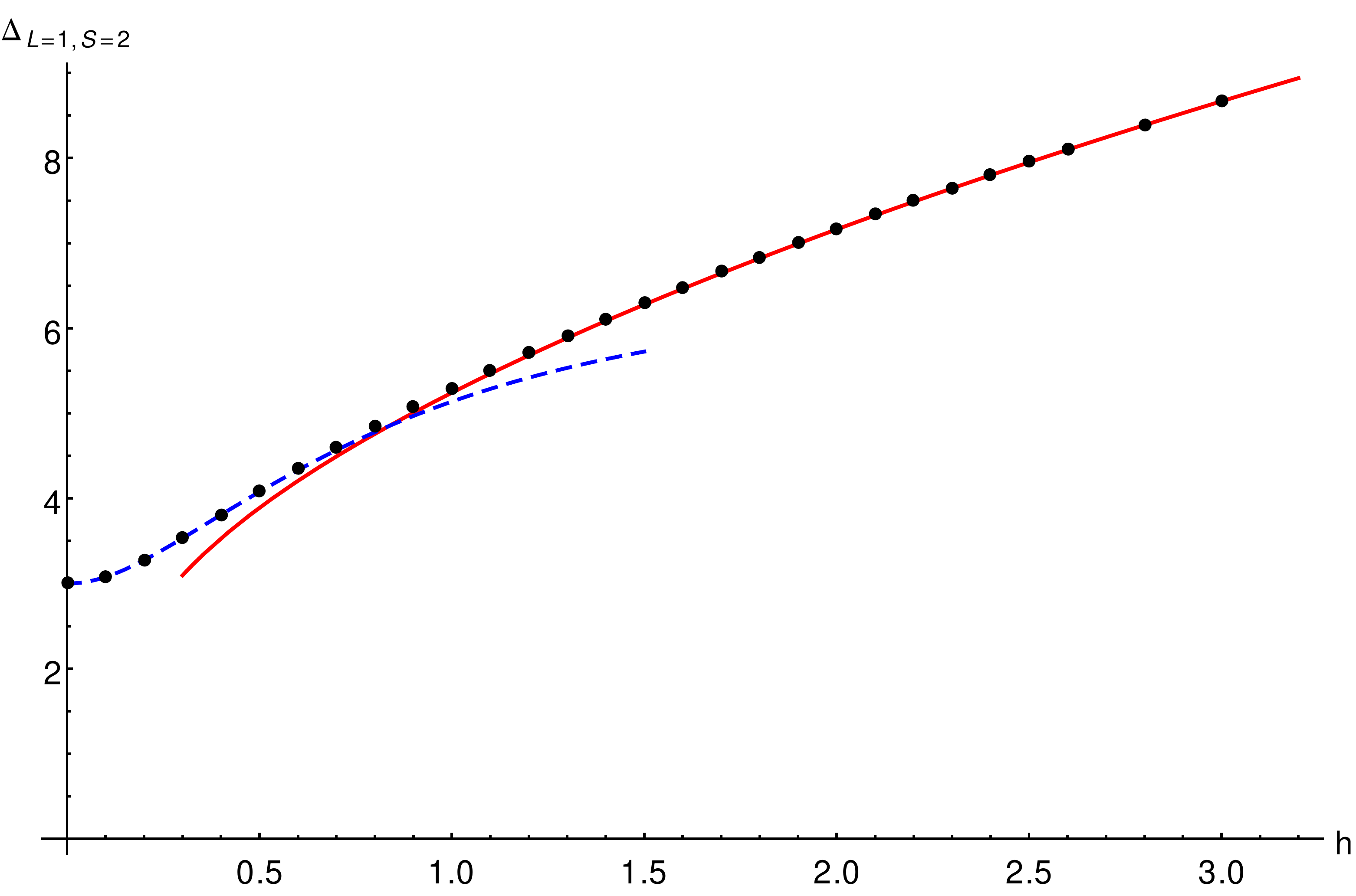}
\caption{Plot of $\Delta_{L=1,S=2}$ as a function of $h$: the dots correspond to the numerical data reported in table \ref{tab:DeltaL1S2} of appendix \ref{app:numericalsl2}, while the solid line and the dashed line represent the strong coupling expansion (\ref{eq:strongpredictionL1S2}) and the weak coupling Pad\'e approximant (\ref{eq:PadeL1S2}), respectively.}
  \label{fig:L1S2}
 \end{figure}

\begin{table}
\begin{center}
\def\arraystretch{1.3}
\begin{tabular}{|c||c|c|c|c|c|c|}
\hline
 $n$ & $\Delta^{(n)}_{\mathrm{fit}}$  & $\Delta_{\mathrm{exact}}$  & $\left|\Delta_{\mathrm{fit}}-\Delta_{\mathrm{exact}}\right|$   \\ 
 \hline
 0 & 1.999999(9) & 2 & 2.1$\times 10^{-8}$  \\
 1 & -0.5000000(1)  & $-\frac{1}{2}$ & 1.2$\times 10^{-8}$ \\
 2 & 1.56254999(9) & $\frac{25}{16}=1.5625$ & 8.9$\times 10^{-9}$  \\
 3 & -2.439979(2) & $\frac{271}{1024}-\frac{9\zeta_3}{4}=-2.43997959\dots$ & 4.4$\times 10^{-7}$ \\
 4 & 12.420858(8) & - & -  \\
  \hline
\end{tabular}
\caption{Strong coupling coefficients for $L=1, S=2$.}
\label{tab:coeffL1S2}
\end{center}
\end{table}

A numerical prediction for the unknown coefficient at order $g^{-5/2}$ is also reported: unfortunately this could not be fixed neither from the exact slope derived in \cite{Gromov:2014eha} nor from the 1-loop results of \cite{Beccaria:2012qd}.
 
Finally, we would like  to remark that also in this case not even the leading order of (\ref{eq:strongpredictionL1S2}) can be predicted correctly by solving the BAEs (\ref{eq:abaL1S2}) in the strong coupling limit:
\beq
e^{\,2\,i\,p-16\,i\,h\cos(p/2)\log\cos(p/2)}=-1\,,\ p=p_1=-p_2\,.
\label{eq:abastrongL1S2}
\eeq
Indeed, the solution of (\ref{eq:abastrongL1S2}) is $p=\sqrt{\frac{\pi}{8h}}+\mathcal{O}\left(\frac{1}{h}\right)$, leading to
\beq
\Delta_{L=1,S=2}^{ABA}=\sqrt{g}+\mathcal{O}(1)\;,
\eeq
which differs from the correct result (\ref{eq:strongpredictionL1S2}) by a factor 2.

To get (\ref{eq:abastrongL1S2}), only the so-called AFS phase \cite{AFS} is needed as leading order contribution of the dressing factor $\sigma_{}$: a similar computation was performed for the first time in \cite{AFS} for the $\mathfrak{su}(2)$ sector of $\mathcal{N}=4$ SYM, where also the $\mathcal{O}(1)$ term was correctly predicted by the leading order BAEs, as noted also in \cite{Rej:2009dk, Arutyunov:2005hd}. 
\section{Spectrum at finite coupling: non-symmetric ${\bf\mathfrak{sl}(2|1)}$ operators}
\label{sec:sl21}
The QSC equations reported in section \ref{sec:review} allow to explore also other sectors, less studied compared to $\mathfrak{sl}(2)$.
In general, the single-trace $\mathfrak{sl}(2|1)$ operators contain fermionic fields $D_+$, $\psi_{4+}$ and $\psi_+^{1\dagger}$ acting on the vacuum $(Y^1Y_4^{\dagger})^L$.
The corresponding BAEs are obtained by setting $K_1=K_2=K_3=0$ in the $\mathfrak{sl}(2)$ grading of the full $Osp(2,2|6)$ BAEs of \cite{Gromov:2008qe}:
{ \small
\beqa
\left(\frac{x_{4,k}^+}{x_{4,k}^-}\right)^L=   \prod_{j=1}^{K_{\bar4}}\frac{u_{4,k}-u_{\bar4,j}-i}{u_{4,k}-u_{\bar4,j}+i} \, \frac{1-\frac{1}{x_{4,k}^+x_{\bar4,j}^-} }{1-\frac{1}{x_{4,k}^-x_{\bar4,j}^+}} \, \sigma(u_{4,k},u_{\bar4,j}) \,  \prod_{j=1}^{K_{4}}\frac{1-\frac{1}{x_{4,k}^+x_{4,j}^-} }{1-\frac{1}{x_{4,k}^-x_{4,j}^+}} \, \sigma(u_{4,k},u_{4,j})\;,
\label{eq:bethe1}\\
\left(\frac{x_{\bar4,k}^+}{x_{\bar4,k}^-}\right)^L=   \prod_{j=1}^{K_{4}}\frac{u_{\bar4,k}-u_{4,j}-i}{u_{\bar4,k}-u_{4,j}+i} \, \frac{1-\frac{1}{x_{\bar4,k}^+x_{4,j}^-} }{1-\frac{1}{x_{\bar4,k}^-x_{4,j}^+}} \, \sigma(u_{\bar4,k},u_{4,j}) \,  \prod_{j=1}^{K_{\bar4}}\frac{1-\frac{1}{x_{\bar4,k}^+x_{\bar4,j}^-} }{1-\frac{1}{x_{\bar4,k}^-x_{\bar4,j}^+}} \, \sigma(u_{\bar4,k},u_{\bar4,j})\;,
\label{eq:bethe2}
\eeqa }
without imposing any particular relation between Bethe roots of type 4 and $\bar4$, except for the ZMC
\beq
\prod_{\alpha=4,\bar4}\prod_{k=1}^{K_\alpha}\frac{x_{\alpha,k}^+}{x_{\alpha,k}^-}=1\,.
\label{eq:zmcsl21}
\eeq
The conformal dimensions in the ABA limit are then given by
\beq
\label{eq:DeltaABAsl21}
\Delta^{ABA}_{\mathfrak{sl}(2|1)} = L + S +\frac{1}{2}\sum_{\alpha=4,\bar4}\sum_{k=1}^{K_{\alpha}}\left(\sqrt{1+16\,h^2\sin^2\left(\frac{p_{\alpha,k}}{2}\right)}-1\right)\,,
\eeq
with $e^{ip_{\alpha,k}}=\frac{x_{\alpha,k}^+}{x_{\alpha,k}^-}\,,\;(\alpha=4,\bar{4}) \,,$ and $S=\frac{K_4+K_{\bar4}}{2}$. The total momenta for each kind (4 or $\bar4$) of particles are
\beq
P_\alpha=\sum_{k=1}^{K_\alpha}p_{\alpha,k} \;,\; (\alpha=4,\bar{4}) \;.
\eeq

Besides the intrinsic physical interest, the QSC equations for non-symmetric states exhibit novel features which are worth to be investigated, for instance the appearance of the non-trivial phase $\sigtw(h)$ in equation (\ref{eq:perioanti}).
In addition, this computation can also be considered as a strong test for the consistency of the QSC in its general form. Indeed, the derivation of the gluing conditions for non symmetric operators in \cite{Bombardelli:2017vhk} was based on an unproven conjecture for the asymptotics of the functions $\tau_i $. Furthermore, our findings confirm that the equations of  \cite{Bombardelli:2017vhk}  form a closed system even without the exact knowledge of the state-dependent function $\mathcal{P}(h)$.

\subsection{The  {\emph L} = 2, {\emph S} = 1 ({\emph K}\textsubscript{4} = {\emph K}\textsubscript{\={4}} = 1) operator}
\label{sec:nonsym}

We start by considering the state in the $\mathfrak{sl}(2|1)$ sector with $L=2$, $K_4=K_{\bar 4}=1$ but different Bethe roots $u_{4,1}=- u_{\bar4,1}=\frac{1}{2 \sqrt{3}}+\mathcal{O}(h^2)$, solution of the ZMC (\ref{eq:zmcsl21}) and the BAEs
\beq
\left(\frac{x_{4,1}^+}{x_{4,1}^-}\right)^2=   \frac{2\,u_{4,1}-i}{2\,u_{4,1}+i} \, \frac{1+\frac{1}{(x_{4,1}^+)^2} }{1+\frac{1}{(x_{4,1}^-)^2}} \, \sigma_{}(u_{4,1},-u_{4,1})\,,
\label{eq:betheL2S1}
\eeq
as a particular case of (\ref{eq:bethe1})--(\ref{eq:bethe2}).
The resulting conformal dimension in the ABA approximation is
\beq
\Delta_{L=2,S=1}^{ABA}= 2+\sqrt{1+16\,h^2\sin^2\left(\frac{p_{4,1}}{2}\right)}\;.
\eeq
For this state $P_4=-P_{\bar4}=\frac{2\pi}{3}+\mathcal{O}(h^2)$, therefore it is one of the simplest example where the total momentum for particles of kind $4$ and $\bar4$ is different from 0 or $\pi$, corresponding to a non-trivial weak coupling value for $\sigtw(h)$: $\sigtw(0)=\frac{2\pi}{3}$. 

\begin{table}
\begin{center}
\def\arraystretch{1.3}
\begin{tabular}{|c||c|c|c|c|c|c|}
\hline
 $n$ & $\Delta^{(n)}_{\mathrm{fit}}$ & $\Delta^{(n)}_{\mathrm{guess}}$ & $|\Delta^{(n)}_{\mathrm{fit}}-\Delta^{(n)}_{\mathrm{guess}}|$ \\ 
 \hline
 0 & 1.999998(9) & 2 & 1.1$\times 10^{-6}$ \\
 1 & -0.49999(9)  & $-\frac{1}{2}$ & 1.4$\times 10^{-6}$ \\
 2 & 1.56250(0)  & $\frac{25}{16}=1.5625$ & 3.5$\times 10^{-7}$ \\
 3 & -2.8149(8) & $-\frac{113}{1024}-\frac{9\zeta_3}{4}=-2.814979595\dots$ & 2.1$\times 10^{-6}$\\
  \hline
\end{tabular}
\caption{Coefficients of (\ref{eq:strongansatz}) for the non-symmetric state with $L=2,\,S=1$, $u_4\neq u_{\bar4}$.}
\label{tab:coeffL2S1}
\end{center}
\end{table}

As in the previous cases, weak coupling expansions of $\Delta$ and the $\bP$'s are necessary as initial input for the iterative procedure.
Since they are not available in the literature, they are computed from scratch adapting the algorithm developed in \cite{Anselmetti:2015mda} and using the symmetric large-$u$ asymptotics (\ref{eq:asymsl2}) for the $\bP$'s but different ansatzs for the large-$x$ expansions of $\bP_5$ and $\bP_6$. The corresponding 8-loop perturbative results are reported in appendix \ref{app:Pweak}. Besides serving as initial input of the numerical algorithm, they may be considered as original findings interesting by their own: in particular, it turns out that $\bP_6(u)=\bP_5(-u)$.

In contrast to the cases discussed previously, in the current case not all the $\bP$'s have a definite parity in $x(u)$. As a direct consequence of this fact a resonance problem appears when solving (\ref{eq:defQai}) in the large-$u$ limit, due to the overlap between some of the exponents of (\ref{eq:Qaiasym}). This problem is overcome following the strategy described in appendix \ref{app:resonance}.

The numerical data in appendix \ref{app:numericalsl21}, table \ref{tab:DeltaL2S1}, are used to predict the strong coupling coefficients reported in table \ref{tab:coeffL2S1}, and allow us to conjecture the following strong coupling expansion for the spectrum of this operator:
\beq
\label{eq:strongnonsym}
\Delta_{L=2,S=1}= 2\,\sqrt{g} -\frac{1}{2} + \frac{25}{16\,\sqrt{g}}-\left(\frac{113}{1024}+\frac{9\zeta_3}{4}\right)\frac{1}{g^{3/2}}+\mathcal{O}\left(\frac{1}{g^{5/2}}\right)\;.
\eeq 
The proposal (\ref{eq:strongnonsym}) is based on the known results for the symmetric operators (\ref{eq:strongprediction}) and (\ref{eq:strongpredictionL1S2}). The differences between $\Delta_{\mathrm{fit}}^{(n)}$ and $\Delta_{\mathrm{guess}}^{(n)}$ are displayed in the last column of table \ref{tab:coeffL2S1}, supporting the correctness of (\ref{eq:strongnonsym}).

Notice that the leading order coefficient in (\ref{eq:strongnonsym}) can be also obtained from the large $h$ limit of (\ref{eq:betheL2S1}):
\beq
e^{\,3\,i\,p-8\,i\,h\cos(p/2)\ln\cos(p/2)}=1\,,\ p=p_{4,1}=-p_{\bar4,1} \;,
\label{eq:strongBAnonsym}
\eeq
which is solved by the ansatz $p=\frac{p_0}{\sqrt{h}}+\frac{p_1}{h}+\dots$ with $p_0=\sqrt{2\pi}$ and $p_1=-\frac{3}{2}$. The resulting ABA prediction for the conformal dimension is then
\beq
\label{eq:DeltaL2S1ABA}
\Delta_{L=2,S=1}^{ABA}=2\,\sqrt{g}-1+\mathcal{O}\left(\frac{1}{\sqrt{g}}\right) \;.
\eeq
Therefore, while there is a mismatch by a factor 2 in the coefficient of the subleading term, surprisingly the leading term in (\ref{eq:strongnonsym}) and (\ref{eq:DeltaL2S1ABA}) are the same. We shall return on this issue in section \ref{sec:L02}.
 
\begin{figure}
\centering
 \includegraphics[width=0.6\textwidth]{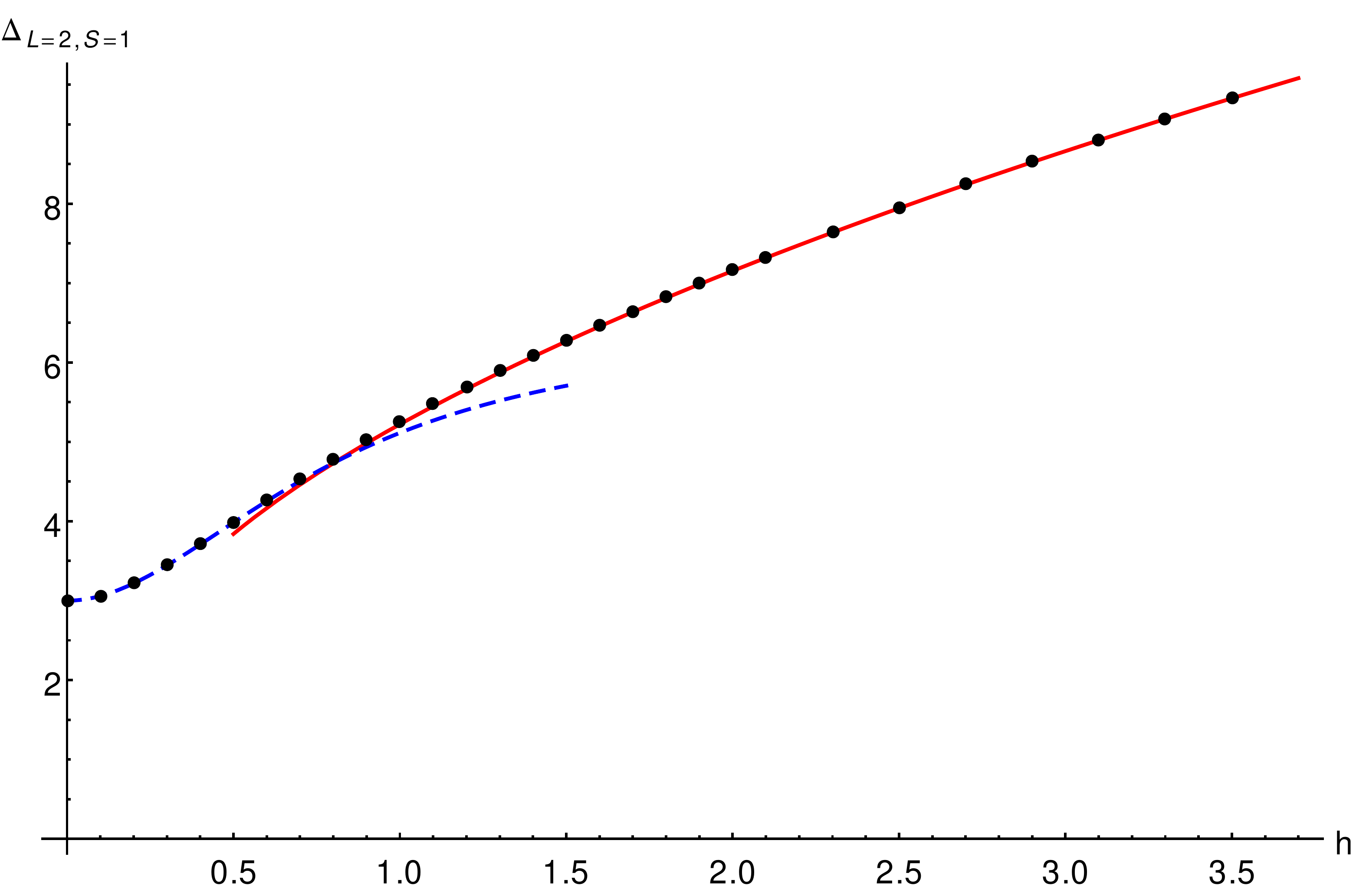}
\caption{Plot of $\Delta_{L=2,S=1}$ as a function of $h$: the dots correspond to the numerical data reported in table \ref{tab:DeltaL2S1} of appendix \ref{app:numericalsl21}, while the solid line and the dashed line represent the strong coupling expansion (\ref{eq:strongnonsym}) and the weak coupling Pad\'e approximant (\ref{eq:Padenonsym}) respectively.}
  \label{fig:nonsym}
 \end{figure}

Finally, in figure \ref{fig:nonsym} the numerical results are compared with the strong coupling expansion (\ref{eq:strongnonsym}) and a diagonal $[4/4]$ Pad\'e approximant of the weak coupling analytic result (\ref{eq:Deltanonsym})
\beq
\label{eq:Padenonsym}
\Delta_{L=2,S=1}^{\mbox{\scriptsize Pad\'e}}=\frac{3+42.5837\,h^2+102.4145 \,h^4}{1+12.1946\,h^2+15.7491\,h^4}\;,
\eeq 
showing nice agreement.

\subsubsection{Computation of $\sigtw$}
\label{sec:calP}

\begin{figure}
\centering
 \includegraphics[width=0.6\textwidth]{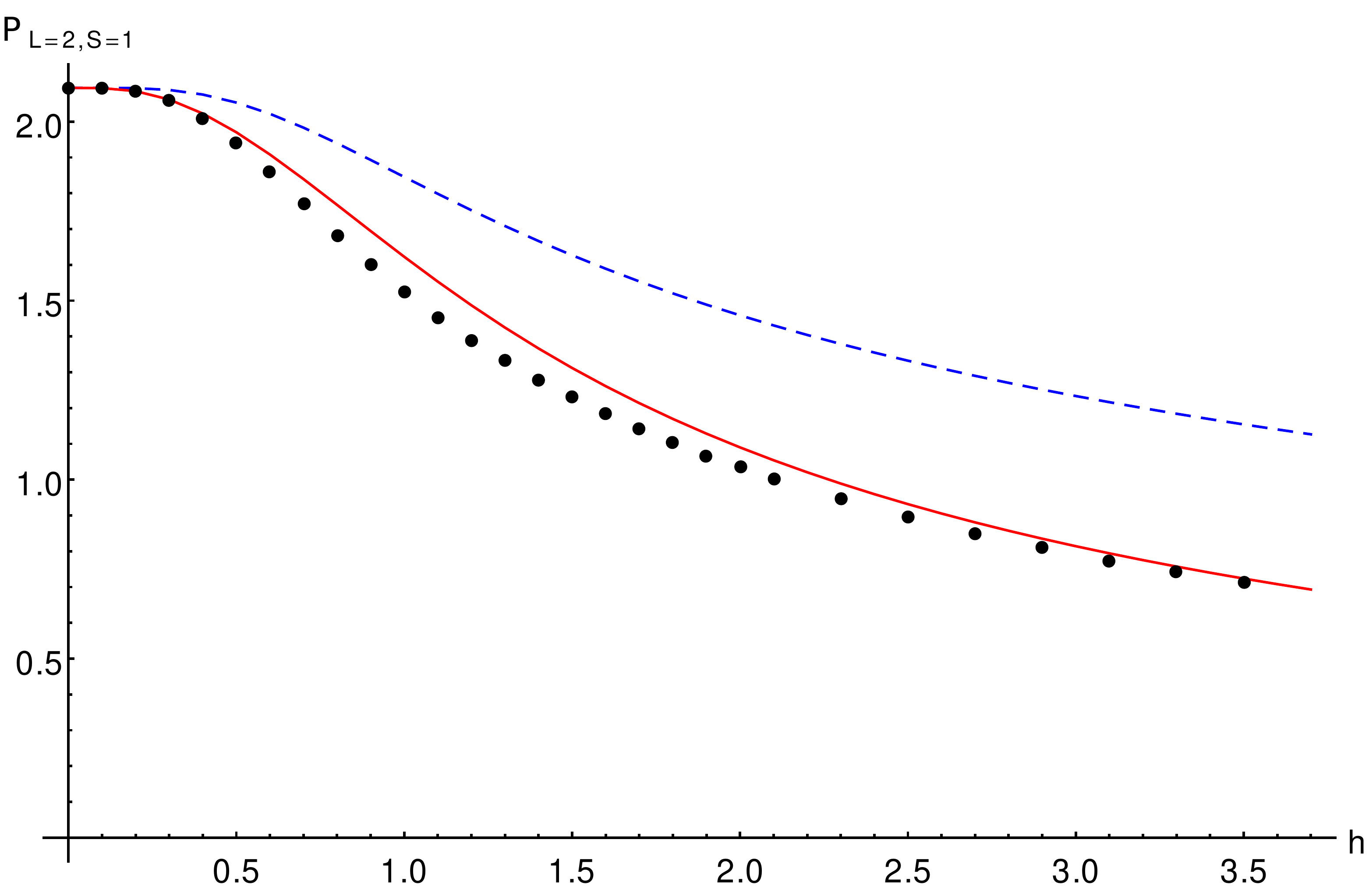}
\caption{Plot of $\sigtw_{L=2,S=1}$ as a function of $h$: the dots correspond to the exact numerical data, the solid line interpolates the values of $\sigtw^{ABA}_{L=2,S=1}(h)$ and the dashed line represents $P^{ABA}_{4,L=2,S=1}(h)$, calculated by (\ref{eq:PABA}) and (\ref{eq:P4ABA}) respectively.}
  \label{fig:calP}
 \end{figure}
The generalisation of the algorithm developed in \cite{Anselmetti:2015mda} to non-symmetric sectors allows us to compute analytically the first five non-trivial coefficients of $\sigtw(h)=-i\log{\frac{\nu_a(0)}{\nu_a(i)}} \;,\; (a=1,\dots,4)$ for the $L=2,S=1$ operator:
\beqa
&&\hspace{-0.5cm}\sigtw_{L=2,S=1}(h)=\frac{2\pi}{3}-\frac{\sqrt{3}\pi^2}{2}\,h^4+\sqrt{3}\left(\frac{\pi^2}{2}+\frac{37\pi^4}{48}-\frac{3\pi^4\Log(2)}{4}-48\,\zeta_3+\frac{7\pi^2\,\zeta_3}{4}+\frac{465\,\zeta_5}{8}\right)h^6\nonumber\\
&&+\sqrt{3}\left(\frac{15\pi^2}{2}-\frac{503\pi^4}{240}-\frac{229\pi^6}{210}-\frac{27\pi^2\Log(2)}{2}+\frac{\pi^4\Log(2)}{8}+\frac{7\pi^6\Log(2)}{8}+27\pi^2\Log^2(2)\right.\nonumber\\
&&+333\,\zeta_3 +\frac{203\pi^2\,\zeta_3}{4}+\frac{\pi^4\,\zeta_3}{32}-567\,\zeta_3\Log(2)-\frac{147\pi^2\,\zeta_3\Log(2)}{2}+\frac{4491\,\zeta_3^2}{16}+\frac{2256\,\zeta_5}{8}\\
&&\left.-\frac{155\pi^2\,\zeta_5}{8}+\frac{1395\,\zeta_5\Log(2)}{2}-\frac{66675\,\zeta_7}{64}+324\,\zeta_{1,-3}+33\pi^2\,\zeta_{1,-3}-360\,\zeta_{1,-5}\right)h^8+\dots\;.\nonumber
\label{eq:ancalP}
\eeqa

In general, the exact expression for $\sigtw(h)$ was first derived in \cite{Bombardelli:2017vhk}. Taking into account the non-trivial monodromy of the logarithm the complete result is
\beqa
\sigtw(h) = n \pi + \frac{1}{4 \pi \, \mathbb{E}(h) } \int_{-2h}^{+2h} dz\, e^{\pi z} \frac{ \log\left( \frac{\tau_4(z)\tilde{\tau}_4(z) }{\tau^1(z)\tilde{\tau}^1(z) } \right) }{\sqrt{(e^{2 \pi z}- e^{4 \pi h} ) \, (e^{ 2 \pi z}- e^{- 4 \pi h} ) } } \;,\; (n\in\mathbb{Z}) \;,
 \label{eq:intePapp2}
\eeqa
where $\mathbb{E}(h)$ is a function of $h$ defined as
\beqa
\mathbb{E}(h) &=& -\frac{1}{2\pi i} \int_{-2h}^{+2h} dz\, \frac{e^{\pi z} }{\sqrt{(e^{2 \pi z}- e^{4 \pi h} ) \, (e^{ 2 \pi z}- e^{- 4 \pi h} ) } } \\
&=& \frac{e^{2 \pi  h}}{2 \pi ^2 i} \left[F\left(\mbox{arcsin}\left(e^{-4 h \pi }\right)\left|e^{8 h \pi }\right.\right)-\mathbb{K}\left(e^{8 h \pi }\right)\right]\;,
\label{eq:Edef}
\eeqa
with $F(z|k^2)$ and $\mathbb{K}(k^2)$ being the incomplete and complete elliptic integral of first kind with modulus $k$, respectively.

The sign ambiguity in $e^{i \sigtw}$ can be lifted by comparison with the leading order of the weak coupling expansion, indeed it corresponds to the $\pm 1$ ambiguity observed in the symmetric sector in \cite{Anselmetti:2015mda}.
In order to compute (\ref{eq:intePapp2}), we need to evaluate
\beq
\textbf{F}=\log\frac{\tau_4 \, \widetilde \tau_4 }{\tau^1 \, \widetilde \tau^1} \;.
\eeq
Using the $\bQ \tau$-system (\ref{eq:QtilTautil}), the quantity $e^{\textbf{F}}$ can be written in terms of the output of our numerical algorithm as
\beq e^{\textbf{F}} = \frac{\sum_{i=1}^4 \tau_4 \, \bQ_{4i} \, \tau^i }{ \sum_{j=1}^4 \tau^1 \, \bQ^{1j} \, \tau_j } = \frac{\sum_{i=1}^4 \bQ_{4i} \, f_4^i }{\sum_{j=1}^4 \bQ^{1j} \, f_j^1 } \label{eq:eq1pap} \; ,
\eeq
where the matrix of functions $f_i^j(u) = \delta_i^j - \tau_i(u) \, \tau^j(u)$ can be computed  as a particular combination of $Q$ functions, as explained in appendix \ref{app:Y}. This allows us to compute $\sigtw_{L=2,S=1}(h)$ non-perturbatively.  

The numerical finite coupling results are displayed in figure \ref{fig:calP} and compared with the ABA expression of the total momentum $P_4$
\beq
P^{ABA}_4(h)=-i\sum_{k=1}^{K_4}\log\frac{x_{4,k}^+}{x_{4,k}^-}\;,
\label{eq:P4ABA}
\eeq
and the ABA approximation \cite{Bombardelli:2017vhk} of $\sigtw(h)$
\beq
\sigtw^{ABA}(h)=n\pi-\frac{1}{4 \pi \, \mathbb{E}(h) } \, \int_{-2 h}^{2 h}\frac{ \log\left(\frac{\mathbb{Q}_4^+(z)}{\mathbb{Q}_4^-(z)}\frac{\mathbb{Q}_{\bar4}^-(z)}{\mathbb{Q}_{\bar4}^+(z)} \right) \, e^{ \pi z} }{\sqrt{ (e^{2 \pi z} -e^{4 \pi h}) \, (e^{2 \pi z} - e^{-4 \pi h} )} } \, dz \;,\; (n\in \mathbb{Z})\;,
\label{eq:PABA} 
\eeq
where $\mathbb{Q}_{\alpha}(u)=\prod_{j=1}^{K_{\alpha}}(u-u_{\alpha,j}) \,,\, (\alpha=4, \bar4)$.
As one can see in figure \ref{fig:calP}, while apparently there is no match between $\sigtw_{L=2,S=1}(h)$ and $P^{ABA}_{4,L=2,S=1}(h)$ except for $h\sim 0$, interestingly the exact and ABA results converge both at small and large $h$. Notice also that, using the solution of (\ref{eq:strongBAnonsym}), it is easy to check that $\sigtw^{ABA}_{L=2,S=1}$ tends to zero at strong coupling. 
In general, for any solution of the ABA (\ref{eq:bethe1})--(\ref{eq:zmcsl21}) with Bethe roots scaling as $h$ at strong coupling, the leading order of $\sigtw^{ABA}$  turns out to be quantized  in integer units of $\pi$.
A natural question is whether the exact formula (\ref{eq:intePapp2}) always reduces to (\ref{eq:PABA}) at strong coupling, and therefore if $\sigtw$ is quantized at strong coupling for a generic state.

\subsection{The {\emph L} = 4, {\emph S} = 1 ({\emph K}\textsubscript{4} = 2, {\emph K}\textsubscript{\={4}} = 0) operator}
\label{sec:L02}

\begin{figure}
\centering
 \includegraphics[width=0.6\textwidth]{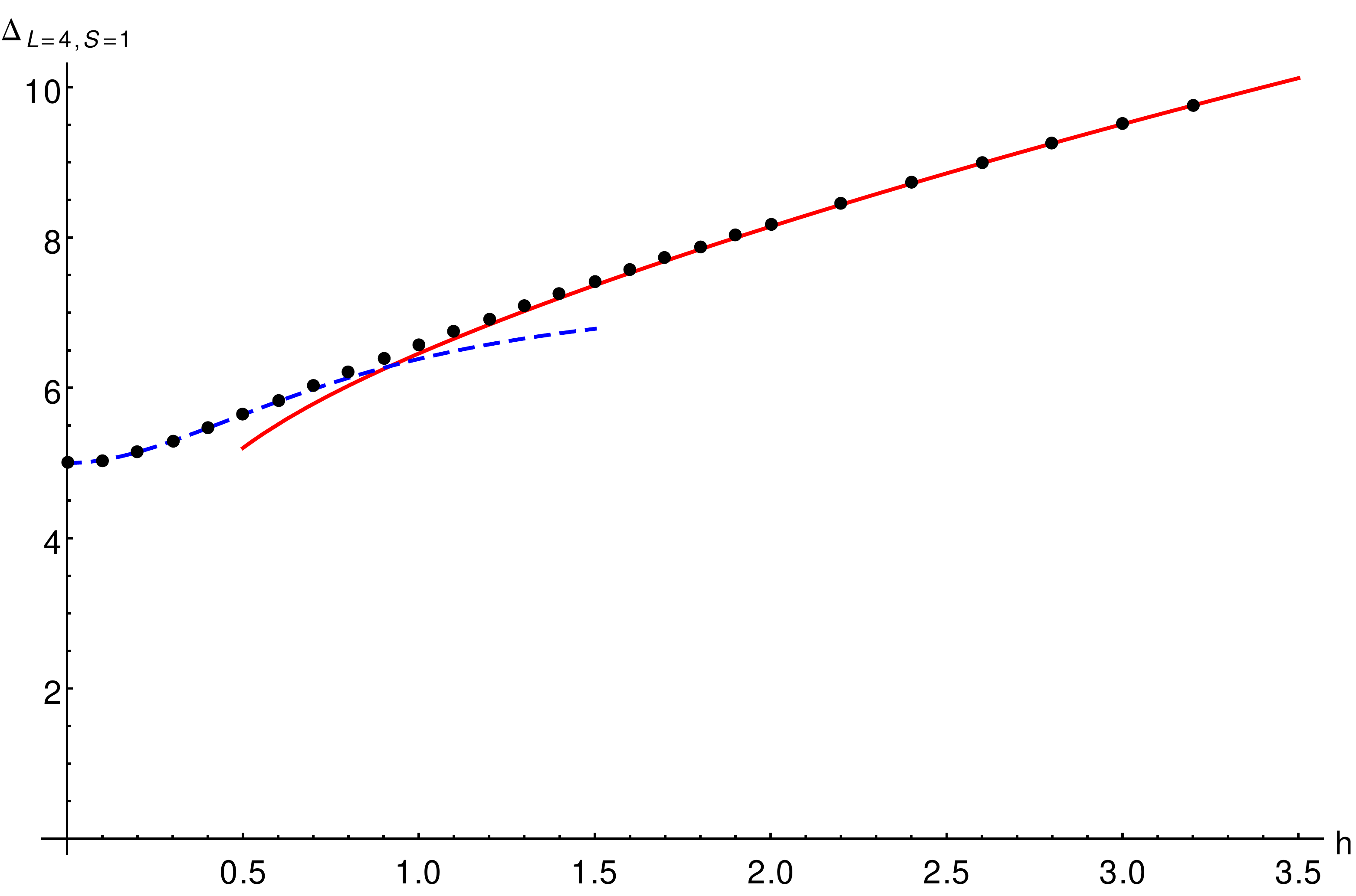}
\caption{Plot of $\Delta_{L=4,S=1}$ as a function of $h$: the dots correspond to the numerical data reported in appendix \ref{app:numericalsl21}, table \ref{tab:DeltaL4S1}, while the solid line and the dashed line represent the strong coupling expansion (\ref{eq:strongL02}) and the weak coupling Pad\'e approximant (\ref{eq:PadeL02}) respectively.}
  \label{fig:L4S1}
 \end{figure}

One of the simplest $\mathfrak{sl}(2|1)$ state with $K_4\neq K_{\bar4}$ is characterised by $L=4$, $K_4=2$, $K_{\bar4}=0$. Then the BAEs (\ref{eq:bethe1})--(\ref{eq:zmcsl21}) reduce to $u_{4,2}=-u_{4,1}$ and
\beq
\left(\frac{x_{4,1}^+}{x_{4,1}^-}\right)^4=\frac{1+\frac{1}{(x_{4,1}^+)^2}}{1+\frac{1}{(x_{4,1}^-)^2}} \, \sigma_{}(u_{4,1},-u_{4,1}) \;.
\label{eq:betheL4S1}
\eeq
In general, for non-symmetric $\mathfrak{sl}(2|1)$ operators the large-$u$ asymptotics of the $\bP$ functions generalise to
\beq
\bP_A\sim\left(\mathcal{A}_1u^{-L},\mathcal{A}_2u^{-L-1},\mathcal{A}_3u^{L+1},\mathcal{A}_4u^{L},\mathcal{A}_5u^{L_0},\mathcal{A}_6u^{-L_0}\right)\,,
\eeq
with $L_0=K_4-K_{\bar4}$.
Differently from the symmetric case, where the following ansatz \cite{Anselmetti:2015mda} for the large-$x$ expansions of $\bP_5$ and $\bP_6$ 
\beq
\bP_5=\bP_6=(x h)^{-L} \left(p_L(u)+\sum_{k=1}^{\infty}c_{0,k}(h)\frac{h^k}{x^k}\right) \;,
\eeq
was used, with $p_L(u)$ being a polynomial of degree $L$, here we use
\beq
\bP_5=(x h)^{-L} \left(p^{(5)}_{L+L_0}(u)+\sum_{k=1}^{\infty}c_{5,k}(h)\frac{h^k}{x^k}\right) \;,\; \bP_6=(x h)^{-L} \left(p^{(6)}_{L-L_0}(u)+\sum_{k=1}^{\infty}c_{6,k}(h)\frac{h^k}{x^k}\right)\;.
\eeq
This is the starting point to generalise the analytic algorithm of \cite{Anselmetti:2015mda} to a state with $L_0\neq 0$, and in particular to compute the 10-loop weak coupling expansions of $\Delta(h)$ and the $\textbf{P}$'s needed as initial input of the program for the state with $L=4$, $S=1$ and $L_0=2$. The resulting expressions are reported in appendix \ref{app:Pweak}, with $\Delta(h)$ matching the ABA result up to $h^8$.

In contrast to the previous case, here all the $\textbf{P}$ functions turn out to have a definite parity in $x(u)$, then no resonance occurs. Moreover, $\sigtw_{L=4,S=1}(h)$ is found analytically to vanish up to $\mathcal{O}(h^{10})$, and this behaviour is confirmed at non-perturbative level by evaluating numerically formula (\ref{eq:intePapp2}) up to $h=3.2$.   
Finally, as already mentioned in section \ref{sec:algorithm}, in this case the inclusion of an extra gluing condition turns out to be necessary to guarantee the convergence of the algorithm.

The numerical results are reported in table \ref{tab:DeltaL4S1} of appendix \ref{app:numericalsl21} and compared, in figure \ref{fig:L4S1}, with a Pad\'e approximant of the weak coupling expansion (\ref{eq:deltaL02})
\beq
\Delta_{L=4,S=1}^{\mbox{\scriptsize Pad\'e}}=\frac{5+50.0103\, h^2+77.8391 \,h^4}{1+9.2021\, h^2+10.6062\, h^4} \;,
\label{eq:PadeL02}
\eeq
and with the conjectured strong coupling asymptotics (see table \ref{tab:coeffL4S1}):
\beq
\Delta_{L=4,S=1}(h)=2\,\sqrt{g}-\frac{1}{2}+\frac{89}{16\,\sqrt{g}}-\left(\frac{5361}{1024}+\frac{9\zeta_3}{4}\right)\frac{1}{g^{3/2}}
+\mathcal{O}\left(\frac{1}{g^{5/2}}\right)\;.
\label{eq:strongL02}
\eeq
Again, the ABA predicts the same leading term as in (\ref{eq:strongL02})
\beq
\Delta_{L=4,S=1}^{ABA}=2\,\sqrt{g}-1+\mathcal{O}\left(\frac{1}{\sqrt{g}}\right)\,.
\eeq
It is quite surprising that, for the non-symmetric $\mathfrak{sl}(2|1)$ states discussed in this paper, the strong coupling limit of $\Delta^{ABA}$ matches the corresponding numerical predictions at leading order.
We do not have a physical explanation for this fact. It may be just an accident, or a specific property that distinguishes between non-symmetric and symmetric operators of the $\mathfrak{sl}(2|1)$ sector in ABJM.

\begin{table}
\begin{center}
\def\arraystretch{1.3}
\begin{tabular}{|c||c|c|c|c|c|c|}
\hline
 $n$ & $\Delta^{(n)}_{\mathrm{fit}}$  & $\Delta^{(n)}_{\mathrm{guess}}$ & $|\Delta^{(n)}_{\mathrm{fit}}-\Delta^{(n)}_{\mathrm{guess}}|$ \\ 
 \hline
 0 & 2.0000000(1)  & 2 & 1.4$\times 10^{-8}$ \\
 1 & -0.4999999(8) &  $-\frac{1}{2}$ & 2.0$\times 10^{-8}$ \\
 2 & 5.5624999(8) &  $\frac{89}{16}=5.5625$ & 1.3$\times 10^{-8}$ \\
 3 & -7.93998(0) &  $-\frac{5361}{1024}-\frac{9\zeta_3}{4}=-7.93997959\dots$ & 2.7$\times 10^{-8}$ \\
  \hline
\end{tabular}

\caption{Strong coupling coefficients for the non-symmetric state with $L=4,\,S=1$, $K_4\neq K_{\bar4}$.}
\label{tab:coeffL4S1}
\end{center}
\end{table}

\section{Conclusions and outlook}

In this paper we presented a numerical method allowing the study of the spectrum of planar ABJ(M) theory at finite coupling, in principle for any operator. The method is based on the Quantum Spectral Curve formulation obtained in \cite{Cavaglia:2014exa,Bombardelli:2017vhk} and on the  numerical algorithm proposed for $\mathcal N=4$ SYM in \cite{Gromov:2015wca}. 
Our results are first of all an important test of the QSC formulation of \cite{Cavaglia:2014exa,Bombardelli:2017vhk}, which itself was based on a long chain of conjectures.\footnote{In fact, this project was carried out in parallel with \cite{Bombardelli:2017vhk} and early numerical results were very important for developing the QSC formulation for the ABJM model.} Besides, attached to this article we provide a simple implementation of the algorithm in {\tt Mathematica},\footnote{For $\mathcal{N}=4$ SYM the algorithm is attached to the arXiv version of \cite{Gromov:2015wca}, see also \cite{Gromov:2017blm}. } which we hope will facilitate future studies of the model.  

The numerical method gives access not only to the spectrum itself but to the full set of Q functions. In the spirit of the Separation of Variables method, the Q functions may play a role not only in encoding the spectrum but also in the description of structure constants and more general observables. Encouraging results in this direction were obtained recently in \cite{Cavaglia:2018lxi} and \cite{Giombi:2018qox}.  

An interesting generalisation of the QSC equations would be to allow for analytic continuation in the spin, similar to what done in $\mathcal{N}=4$ SYM in \cite{Gromov:2014eha,QCDPomeron,Gromov:2015vua,Alfimov:2018cms}, where this allows to reach a BFKL regime, relevant for high-energy scattering, where the theory is similar to QCD. It would be interesting to investigate whether a similar regime exists also for the ABJM model or whether there are qualitative differences. This in turn could help reveal new properties of the spectrum and amplitudes, see  \cite{Beccaria:2009ny}. 
 Allowing for complex spin would require a modification of the algorithm presented here, in particular a change in the gluing conditions. 
 For ABJM theory the first steps in this direction were taken in  \cite{Gromov:2014eha} and \cite{Lee:2017mhh}. 

 Another interesting, almost completely unexplored  problem, is the study of the analytic dependence of the spectrum on the coupling constant. The spectrum has branch points in the complex domain, whose nature can be investigated efficiently numerically. We plan to report on this problem soon  \cite{ComplexToappear}.

It would also be interesting to extend the ABJM QSC to the twisted case, in particular in view of the interest of the 3D integrable fishnet model obtained as a double scaling limit of twisted ABJM theory \cite{Caetano:2016ydc,Mamroud:2017uyz}. Together with the recently much studied 4D fishnet model \cite{Gurdogan:2015csr,Caetano:2016ydc,Gromov:2017cja,Kazakov:2018ugh}, this  non-supersymmetric model, which has an explicit Lagrangian description, allows for a direct  all-loop connection between integrable spin chains and Feynman diagrams and could be very useful to develop the integrability approach for observables beyond the spectrum.

The integrable description of cusped Wilson lines is also naturally a very important open problem for ABJM theory, which could be solved  adapting the QSC method as done in \cite{QSCCusp}. This in particular would allow for many strong tests of the conjecture for $h(\lambda)$ due to the existence of independent localisation results (see e.g. \cite{Bianchi:2014laa,Bianchi:2018bke,Bianchi:2018scb} for recent results)

Finally, in order to make the numerical QSC method universally applicable to any operator, it would be important to develop a systematic weak coupling algorithm covering the full $Osp(2,2|6)$ spectrum, in analogy to the work started in \cite{Marboe:2017dmb} where a fully automatic method of weak coupling expansion for any operator in $\mathcal{N}=4$ SYM was described. This would be very useful since the numerical algorithm requires rather precise initial data for the iterative procedure in order to converge on a given operator. 
It is also worth noticing that the study of the finite coupling regime with numerical methods requires extensive  computational time and power. The situation worsens as the most interesting regimes --- the strong coupling limit and close to branch points for complex coupling  \cite{ComplexToappear} --- are approached. 
Therefore, an optimisation appears to be very desirable to pursue the numerical analysis at a more satisfactory level.

\section*{Acknowledgments}

We thank Mikhail Alfimov, Lorenzo Bianchi, Sergei Frolov, David Grabner, Nikolay Gromov,  Julius, Christian Marboe, Fedor Levkovich-Maslyuk, Stefano Negro and Dmytro Volin for interesting discussions and suggestions. We are especially grateful to Nikolay Gromov, Fedor Levkovich-Maslyuk and Grigory Sizov for sharing the {\tt Mathematica} code for $\mathcal{N}=4$ SYM at an early stage of the project.

This project was partially supported by the INFN project SFT and the EU network GATIS+. 
For most of the duration of this project AC was supported by a postdoctoral fellowship from the University of Turin, and is presently supported by King's College London and the STFC grant (ST/P000258/1).

\appendix

\section{Symmetries of QSC equations and gauge fixing}
\label{app:sym}

\subsection{Symmetry acting on the coefficients $\{ c_{A, n} \}_{n \geq 0}$ and its gauge fixing}
In this appendix we describe the symmetry of the QSC equations which acts directly on the coefficients $\left\{c_{A, n} \right\}_{n\geq 0}$. 
It is easy to verify that the transformation
\beq
\nu_a \rightarrow R_a^b \, \nu_b \;,\; Q_{a|i} \rightarrow R_a^b \, Q_{b|i} \;,
\bP_{ab} \rightarrow ( R\, \bP\, R^T )_{ab} \;,
\label{eq:Rsym}
\eeq
where  $R_a^b$ is a constant unit-determinant matrix, leaves invariant the algebraic form of all the equations presented in section 1, and does not alter the analytic structure of any function involved. 
 In addition, to preserve the asymptotics of $\bP_{ab}$ and $Q_{a|i}$, $R_a^b$ should have the following form
\beq
R_a^b = \left(
\begin{array}{cccc}
 {r_1} & 0 & 0 & 0 \\
 {r_4} & {r_2} & {r_3} & 0 \\
{r_6} & 0 & {r_5} & 0 \\
 {r_9} & {r_7} & {r_8} & \frac{1}{{r_1} {r_2} {r_5}} \\
\end{array}
\right) \;,
\eeq
where we used the fact that the charges are ordered as in (\ref{eq:asyorder}), and $r_i \in \mathbb{R}$ for the reality of the coefficients $\left\{ c_{A, n} \right\}_{n\geq 0}$. Choosing suitably $\left\{ r_1^2 \, ,\, r_2 \,,\, r_5 \right\}$, the transformation (\ref{eq:Rsym}) can be used to fix the values of $\mA_1$, $\mA_2$, $\mA_5$,\footnote{Then notice that $\mA_3$, $\mA_4$, $\mA_6$ are fixed in terms of the charges by the relations (\ref{eq:AAasy}).} in (\ref{eq:Pasy}):
\beq
\mA_1 = a_1 \;,\; \mA_2 = a_2 \;,\; \mA_5 = a_5 \;, \label{eq:aiGF}
\eeq 
where $a_i$ are arbitrary constants. 
The remaining six-parameter freedom can be used to enforce the following additional gauge fixing conditions:
\beqa\label{eq:gauge}
&&c_{5 , M_1 - M_5 } = 0 \;,\; c_{5 , M_2 - M_5 } = 0 \;,\nn\\
&&c_{6 , M_1 - M_6 }  = 0 \;,\; c_{6 , M_2 - M_6 } = 0 \;, \nn\\
 && c_{1 , M_2 - M_1 } = 0 \;,\; c_{3 , M_1 - M_3 } = 0 \;.
\eeqa
Equations (\ref{eq:aiGF}) and (\ref{eq:gauge}), for a given choice of $a_i$, define the reduced space of parameters in which the numerical algorithm operates.\footnote{The particular choice of $a_i$ is irrelevant in principle. However, practically this choice is important for the convergence of the algorithm.  As a rule of thumb, we should choose $a_i$, in such a way that the coefficients $\left\{c_{A, n} \right\}_{n\geq 0}$ are roughly of the same size for any $A$ at any fixed value of $n$. }

\subsection{Residual symmetry acting on $\bQ_{ij}$ }

There exists a further algebraic symmetry of the QSC equations: 
\beq\label{eq:Gtrans}
Q_{a|i}\rightarrow Q_{a|l } \, G^l_i \;,\; \bQ_{ij} \rightarrow  G_i^k \, \bQ_{kl} \, G_j^l \;,
\eeq 
where $G_i^j$ is a constant matrix satisfying $G^j_i \, \kappa_{jk} \, G^k_l=\kappa_{il}$. The large-$u$ asymptotics of the QSC is preserved if $G_i^j$ is  lower-triangular. 

 The condition of pure asymptotics, which is always enforced in our algorithm, breaks this symmetry almost completely, since requiring that $Q_{a|i}(u)$ has large-$u$ expansion of the form (\ref{eq:pure}) forbids a generic mixing between different columns of  this matrix. 
 However, for physical operators it is always true that the  asymptotics of the second and third column of $Q_{a|i}(u)$ differ by an integer (see \cite{Bombardelli:2017vhk}):
\beq\label{eq;defN0}
\hat{\mathcal{N}}_2 - \hat{\mathcal{N}}_3 \equiv 2 S +1 \in \mathbb{N}^+  ,
\eeq
 therefore, a mixing between these two particular columns does not spoil the assumption of a \emph{pure} asymptotic expansion. This implies that the equations used in the numerical algorithm are invariant under a one-parameter family of symmetries, given by (\ref{eq:Gtrans}) with $G_i^j$ taking the form:
\beq\label{eq:Gsym}
G_i^{\; l} = \left(\begin{array}{cccc} 
1&0&0&0 \\
0 & 1 & 0 & 0 \\
0 & f & 1 & 0 \\
0 & 0 & 0 & 1
\end{array} \right),
\eeq
where $f \in \mathbb{R}$ is an arbitrary real parameter\footnote{Notice also that this transformation acts on the $\bQ$ functions as follows: $\bQ_1 \rightarrow \bQ_1 + f \bQ_2$, $\bQ_3 \rightarrow \bQ_3 + f \bQ_4$, leaving the other $\bQ$ functions unchanged. This map preserves the form of the gluing conditions, with exactly the same coefficients.}. 
As we discuss below, the presence of this zero mode in our equations produces a singularity in the linear system used to compute the large-$u$ expansion of $Q_{a|i}$, due to the ``resonance'' between two solutions. The careful treatment of this problem is discussed in detail in appendix \ref{app:resonance}. 

 However, this issue does not arise for all states corresponding to $\bP$ functions with definite parity in $u$, and additionally have integer (as opposed to only half-integer) spin. For these states, we can consistently demand that the large-$u$ expansion (\ref{eq:pure}) goes in even powers of $1/u$ only. This requirement forbids the mixing described by the transformation (\ref{eq:Gsym}), so that in this case the resonance problem discussed below does not occur. 

\subsection{The resonance problem}
\label{app:resonance}

Let us introduce the following notation
\beq\label{eq:defN0}
N_0 \equiv \hat{\mathcal{N}}_2 - \hat{\mathcal{N}}_3  = 2 S +1 \in \mathbb{N}^+ \;.
\eeq
The presence of the symmetry (\ref{eq:Gsym}) implies that the computation of the coefficients $\left\lbrace B_{(a|i), n} \right\rbrace_{n\geq 0}$ is ambiguous for the column $i=2$ and $n \geq N_0$. 
 To be more explicit, let us consider the large $u$ expansion of (\ref{eq:defQai}). At order $1/u^n$, we obtain a linear system of equations of the form:
\beq
\left( M^{(i), n} \right) \, \left(\begin{array}{c} B_{(1|i), n} \\  B_{(2|i), n} \\
 B_{(3|i), n} \\
  B_{(4|i), n} \end{array}\right) = V_n \;,
\eeq
where $M^{(i), n}$ and $V^{(i), n}$ are respectively a $4 \times 4$ matrix and a $4$-vector which contain the coefficients $\left\{c_{A, n} \right\}_{n\geq 0}$ and $\left\{ B_{(a|i), m} \right\}_{m\geq 0}$ for $m < n$. Solving the system order by order, the matrix of coefficients appearing at order $n$ can be fully determined in terms of the coefficients ${\left\{c_{A, j} \right\}_{j=1}^{N_0-1}}$ and the charges, while the vector $V_n$ also depends linearly on the six coefficients $c_{A, N_0}$ with $A=1, \dots, 6$.  

The concrete manifestation of the zero mode is that at the critical level we have 
$$\text{det}(M^{(i=2), n=N_0}) = 0 \;,\; \text{rank}\left( M^{(i=2), n=N_0} \right) = 3 \;. $$ 
This means that system has either zero or a one-parameter family of solutions, depending on whether a certain constraint is met on the vector of coefficients $V^{(i=2), (n=N_0)}$. This constraint gives a linear equation to be satisfied by the coefficients $\left\{c_{A, N_0}\right\}_{A=1}^6$. 

A convenient way to impose this condition is to rewrite the set of equations appearing at $n=N_0$
 as a linear system for a different set of unknowns. Indeed, the linear system:
\beq\label{eq:criticallevel}
\left( M^{(i=2), N_0} \right) \, \left(\begin{array}{c} B_{(1|2), N_0} \\  B_{(2|2), N_0} \\
 B_{(3|2), N_0} \\
  B_{(4|2), N_0} \end{array}\right) = V_{N_0} \;,
\eeq 
can always be rewritten as\footnote{This is due to the fact that the coefficients $c_{A, N_0}$ appear linearly in $V_{N_0}$ and do not appear in $M^{(i), N_0}$.}
\beq\label{eq:newlinear}
\left( \hat{M}^{(i=2), N_0} \right) \, \left(\begin{array}{c} c_{A_0, N_0} \\  B_{(2|2), N_0} \\
 B_{(3|2), N_0} \\
  B_{(4|2), N_0} \end{array}\right) = \hat{V}_{N_0} \;,
\eeq 
 where $A_0$ is an arbitrarily chosen index $1 \leq A_0 \leq 6$. Above,  $\hat{M}^{(i=2), N_0}$ is a new $4 \times 4$ matrix of coefficients and $\hat{V}_{N_0}$ a new $4$-vector. In particular, $\hat{M}^{(i=2), N_0}$ depends on the coefficients $\left\{c_{A, n}\right\}_{n=1}^{N_0-1}$ while $\hat{V}_{N_0}$ additionally depends on $\left\{ c_{A, N_0} \right\}_{A \neq A_0} \cup \left\{ B_{(2|1), N_0} \right\}$. 
 Choosing $A_0$ appropriately, the matrix of coefficients in (\ref{eq:newlinear})  will have non-zero determinant, so that the system can be solved unambiguously. 
 For the operator considered in section \ref{sec:nonsym}, the choice $A_0=2$ ensured that the system was non-singular.\footnote{In every concrete case, it is not difficult to compute $\hat{M}^{(i), n}$ analytically for $n \leq N_0$ and verify this explicitly.} It is also a particularly convenient choice since the coefficient $c_{2, n}$ is not involved in any of the gauge fixing conditions (\ref{eq:gauge}),  and it is therefore clear that the latter do not clash with equation (\ref{eq:newlinear}).  
 Notice that, as expected, we still have a one-parameter family of solutions depending on the unfixed value of $B_{(1, 2), N_0}$. This coefficient is truly arbitrary, and we can set it to zero using the symmetry (\ref{eq:Gsym}). This removes all residual ambiguities from the solution.   

In summary, the proposed procedure is to:
\begin{itemize}
\item Set $B_{(1|2), N_0}=0$. This completely removes the redundancy (\ref{eq:Gsym}). 
 \item Exclude $c_{A_0 = 2, N_0}$ from the set of variational parameters. This coefficient is not varied freely but instead is computed from the linear system appearing at level $n=N_0$, rewritten as  $(\ref{eq:newlinear})$.
\end{itemize} 

\section{Computation of Y\textsubscript{1,0} and other useful quantities}\label{app:Y}

Let us briefly summarise how $\textbf{Y}_{1,0}$ can be computed in terms of the $\bf P$ and $\bf Q$ functions (for the  context necessary to understand the following technical details see appendix A in \cite{Bombardelli:2017vhk}).   
We start from the definition of $\textbf{Y}_{1,0}$ in terms of the $T$ functions 
\beq
\label{eqn:Y vs T}
\textbf{Y}_{1,0}^\alpha = \frac{T_{1,1}T_{1,-1}^\beta}{T_{2,0}^\alpha T_{0,0}^\beta} \;,\; (\alpha,\beta\in\left\lbrace I,II \right\rbrace \;;\; \alpha\neq\beta) \;,
\eeq
where the indexes $\alpha$ and $\beta$ label the two different wings of the T-diagram of ABJM.

Then, using the Hirota equation 
\beq
\bigl(T_{1,0}^\alpha \bigr)^+ \bigl( T_{1,0}^\beta \bigr)^-=T_{2,0}^\alpha T_{0,0}^\beta+T_{1,1}T_{1,-1}^\beta\;,
\eeq
equation~\eqref{eqn:Y vs T} can be recast into
\beq
\frac{1}{\textbf{Y}_{1,0}^\alpha} = \frac{\bigl(T_{1,0}^\alpha \bigr)^+ \bigl( T_{1,0}^\beta \bigr)^- }{T_{1,1}T_{1,-1}^\beta}-1 \;.
\eeq
$T$ functions are affected by gauge ambiguities. There is a convenient gauge choice where the $T$ functions --- denoted as $\mathbb{T}$ --- appearing in~\eqref{eqn:Y vs T} admit a nice representation in terms of the building blocks of the $\textbf{P}\nu$-system
\beq
\mathbb{T}_{1,-1} = 1 \;,\; \mathbb{T}_{1,1} = \textbf{P}_1^+ \textbf{P}_2^- -\textbf{P}_2^+ \textbf{P}_1^- \;,\; \mathbb{T}_{1,0} = \sqrt{\tilde{\nu}_1\tilde{\nu}^4} \;.
\eeq
Considering that both the $Y$- and $T$-systems are naturally defined on the mirror Riemann section and $\nu_1\nu^4$ is $i$-periodic on this section, $\mathbb{T}_{1,0}^\pm$ can be expressed as
\beq
\label{eqn:T10}
\mathbb{T}_{1,0}^+ = \sqrt{(\nu_1 \nu^4)^{[+3]}} \;,\; \mathbb{T}_{1,0}^- = \sqrt{(\nu_1 \nu^4)^-} \;.
\eeq
Inverting the definition of $\tau$ in (\ref{eq:Qijdef}):
\beq
\nu_a = -Q_{a|i}^- \tau^i \;,\; \nu^a = \bigl( Q^{a|i} \bigr)^- \tau_i \;,\; (a,i = 1,\dots,4) \;,
\eeq
we can write
\beq
\nu_1 \, \nu^4 = -Q_{1|i}^- \, \bigl( Q^{4|j} \bigr)^- \, \tau_j \, \tau^i =  Q_{1|i}^- \, \bigl( Q^{4|j} \bigr)^- \, f^i_j \;,
\eeq
using $f_i^j(u) = \delta_i^j - \tau_i(u) \, \tau^j(u)$.

Let us now show how the functions $f_i^j(u)$ can be expressed in terms of Q functions. 
 To do this we introduce a matrix $\Omega_i^j(u)$ which relates $Q_{a|i}$ to its  conjugate $\overline{Q}_{a|i}(u) = \left( Q_{a|i}(u^*) \right)^*$ (see \cite{Bombardelli:2017vhk}):
\beq\label{eq:defOme}
\overline{Q}_{a|i}(u) = Q_{a|j}(u) \; ( \Omega^j_i (u) )^+\; ,
\eeq
where $\Omega_i^j(u+ i) = \Omega_i^j(u)$ due to the fact that (\ref{eq:defQai}) is a real equation. 
As explained in \cite{Bombardelli:2017vhk} this matrix can be used to construct a constant gluing matrix
\beq
\label{eq:defL}
\mathcal{L}_i^j = ( f(u)\, {\Omega}^{-1}(u) )_i^j \;,
\eeq
which can be fixed explicitly to the form 
\beq
\label{eq:calL}
\mathcal{L}_i^j = \left(\begin{array}{cccc}
\frac{ e^{i \pi \hat{M}_1 }}{\cos( \pi \hat{M}_1 ) } & 0& 0 &  \delta_1\\
0 & 1 & 0 & 0 \\
0 & 0 & 1 & 0 \\
\delta_2  & 0 & 0 &  \frac{  e^{-i \pi \hat{M}_1 }}{\cos( \pi \hat{M}_1 ) } 
\end{array}\right) \; ,
\eeq
where the constants $\delta_i$ are the same appearing in the gluing conditions and can be obtained as an output of the numerical algorithm. 
From (\ref{eq:defL}), we have
\beq
\label{eq:fij}
f_i^j(u) = \mathcal{L}_i^k \, \Omega_k^j(u) = \mathcal{L}_i^k \, (\overline{Q_{a|k}} )^{-} \, (Q^{a|j} )^- \; ,
\eeq
so that the product $\nu_1\nu^4$ can be expressed in terms of the $Q$ functions as
\beq
\mathbb{T}_{1,0}^+ = \sqrt{ \left( Q^{4|i} \right)^{[+2]} \mathcal{L}_i^{\hspace{1mm}k} \; \overline{ Q_{1|k}^{[-2]} } } \;,\; \mathbb{T}_{1,0}^- =  \sqrt{ \left( Q^{4|i} \right)^{[-2]} \mathcal{L}_i^{\hspace{1mm}k} \; \overline{ Q_{1|k}^{[+2]} } }\;.
\eeq
In this way, $\textbf{Y}_{1,0}$ is computed from the $\textbf{P}$ and $\textbf{Q}$ functions only. 

Finally, using (\ref{eq:fij}), formula (\ref{eq:eq1pap}) for the integrand of the integral formula for $\mathcal{P}(h)$ can be recast into 
\beq
e^{\textbf{F}} = \frac{{\mathcal{L}_4^k \, (\overline{Q_{a|k}} )^{-} \, \bP^{ab} \, Q_{b|4}^- }}{{(Q^{a|1})^- \, \bP_{ab} \, (\overline{Q^{b|j}} )^- \, (\mathcal{L}^{-1} )_j^1}} \;,
\label{eq:F}
\eeq
where the r.h.s. is expressed in terms of the Q functions computed by our numerical algorithm.

\section{Explicit $\bP$ functions and ${\bf\Delta}$({\emph h}) at weak coupling}
\label{app:Pweak}

Here below we report the perturbative results used as input data of the numerical algorithm in the case of the $\mathfrak{sl}(2|1)$ operator with $L=2$, $K_4=K_{\bar4}=1$, {\it i.e.}  the analytic solutions of $\Delta_{L=2,S=1}(h)$ and the related $\bP$ functions up to the fourth non-trivial order at small $h$:
{\small
\beqa
\label{eq:Deltanonsym}
&&\Delta_{L=2,S=1}=3+6\,h^2-18\,h^4+\left(54+3\,\pi^2+\frac{17\,\pi^4}{40}\right) h^6 \\
&&+\left(-423\,\zeta_3+\frac{141\,\pi^2\,\zeta_3}{4}-\frac{1485\,\zeta_5}{4}+486-78\,\pi^2-\frac{19\,\pi^4}{8}-\frac{325\,\pi^6}{504}+54\,\pi^2\,\log (2)\right)h^8+\dots\,,\nonumber
\eeqa}
{\small\beqa
\label{eq:Pnonsym1}
&&\bP_1(u)=\frac{1}{u^2}+\frac{2 h^2}{u^4}+h^4
   \left(\frac{5}{u^6}+\frac{3}{u^4}\right)+h^6 \left(\frac{14}{u^8}+\frac{12}{u^6}+\frac{12 \zeta_3}{u^4}-\frac{18}{u^4}\right)+\dots\,,\\
&&\bP_2(u)=\frac{h^2}{u^3}+\frac{3 h^4}{u^5}+\frac{9 h^6}{u^7}+h^8 \left(\frac{28}{u^9}+\frac{12 \zeta_3}{u^5}\right)+\dots\,,\nonumber\\
&&\bP_3(u)=\frac{18}{5 u}+4u-\frac{28 u^3}{5}+h^2 \left(-\frac{136 u^3}{5}+\frac{21}{5 u^3}+\frac{304 u}{5}+\frac{506}{5 u}\right)+h^4 \left(\frac{9}{u^5}-\frac{119}{300} \pi ^4 u^3\right.\nonumber\\
&&\left.-\frac{14 \pi ^2 u^3}{5}+\frac{732 u^3}{5}+\frac{666}{5 u^3}+\frac{72 \zeta_3}{u}+\frac{17 \pi ^4 u}{60}+2 \pi ^2 u+\frac{144 u}{5}-\frac{17 \pi ^4}{100 u}+\frac{4 \pi ^2}{5 u}+\frac{4218}{5 u}\right)\nonumber\\
&&+h^6\left(\frac{117}{5 u^7}+\frac{1506}{5 u^5}+\frac{693 u^3 \zeta_5}{2}-\frac{329}{10} \pi ^2 u^3 \zeta_3+\frac{1974 u^3 \zeta_3}{5}+\frac{576 \zeta_3}{5 u^3}+\frac{65 \pi ^6
   u^3}{108}\right.\nonumber\\
&&-\frac{241 \pi ^4 u^3}{60}+\frac{144 \pi ^2 u^3}{5}-\frac{5328 u^3}{5}-\frac{17 \pi ^4}{50 u^3}+\frac{8 \pi ^2}{5 u^3}+\frac{6378}{5 u^3}-\frac{252}{5} \pi ^2 u^3 \log
   (2)-\frac{495 u \zeta_5}{2}\nonumber\\
   &&+\frac{47}{2} \pi ^2 u \zeta_5-\frac{1266 u \zeta_3}{5}-\frac{1143 \zeta_5}{2 u}-\frac{141 \pi ^2 \zeta_3}{10 u}+\frac{7956 \zeta_3}{5 u}-\frac{325
   \pi ^6 u}{756}+\frac{1289 \pi ^4 u}{100}\nonumber\\
   &&\left.-\frac{378 \pi ^2 u}{5}-72 u+\frac{325 \pi ^6}{1008 u}+\frac{163 \pi ^4}{40 u}+\frac{146 \pi ^2}{5 u}+\frac{2622}{5 u}+36 \pi ^2 u \log
   (2)+\frac{72 \pi ^2 \log (2)}{5 u}\right)+\dots\,,\nonumber\\
&&\bP_4(u)=3-3u^2+h^2\left(\frac{18}{5u^2}+76-\frac{168u^2}{5}\right)+h^4\left(\frac{39}{5u^4}+\frac{476}{5u^2}+\frac{2506}{5}-\frac{81u^2}{5}\right)\nonumber\\
&&+h^6\left(\frac{102}{5 u^6}+\frac{1088}{5 u^4}+\frac{36 \zeta_3}{u^2}-\frac{119}{50} \pi ^4 u^2-\frac{84 \pi ^2 u^2}{5}+\frac{1242 u^2}{5}-\frac{17 \pi ^4}{100 u^2}+\frac{4 \pi ^2}{5
   u^2}+\frac{3696}{5 u^2}\right.\nonumber\\
&&\left.+180 \zeta_3+\frac{661 \pi ^4}{120}+11 \pi ^2-\frac{1308}{5}\right)+\dots\,,\nonumber\\
&&\bP_5(u)=2-\frac{\sqrt{3}}{u}+h^2 \left(-\frac{\sqrt{3}}{u^3}-\frac{11 \sqrt{3}}{u}+7\right)+h^4 \left(-\frac{2 \sqrt{3}}{u^5}+\frac{1}{u^4}-\frac{11 \sqrt{3}}{u^3}-\frac{3\sqrt{3}}{u}-15\right)\nonumber\\
&&+h^6 \left(-\frac{5 \sqrt{3}}{u^7}+\frac{4}{u^6}-\frac{23 \sqrt{3}}{u^5}+\frac{3}{u^4}-\frac{3 \sqrt{3}}{u^3}-\frac{24 \sqrt{3} \zeta_3}{u}-\frac{209 \pi ^4}{80 \sqrt{3} u}-\frac{3 \sqrt{3}
   \pi ^2}{2 u}+\frac{81 \sqrt{3}}{u}\right.\nonumber\\
&&\left.+\frac{119 \pi ^4}{240}+\frac{7 \pi ^2}{2}+27\right)+\dots\,,\nonumber\\
&&\bP_6(u)=\bP_5(-u)\,.\nonumber
\label{eq:Pnonsymlast}
\eeqa}
The coefficients $c_{A,i}(h)\;,\; (A=1,\dots,6 \;;\; i=1,\dots, 6)$ of the large-$x$ expansion of the $\bP$'s used as initial condition of the program at weak coupling (for $h$ up to 0.3) then read
{\small\beq
c_{A,i}=\left(
\begin{array}{cccccc}
 0 & 3 h^2 & 0 & 0 & 0 & 0 \\
 0 & 0 & 0 & 0 & 0 & 0 \\
 0 & 4h+44h^3+ h^5\left(-\frac{264}{5}+2 \pi ^2+\frac{17 \pi ^4}{60}\right) & 0 & \frac{18}{5 h}+\frac{526 h}{5} & 0 & \frac{3}{5 h}+32 h \\
 0 & 0 & 0 & 0 & 0 & 0 \\
-\frac{\sqrt{3}}{h}-11 \sqrt{3} h -3 \sqrt{3} h^3 & 0 & 0 & 1 & 0 & 0 \\
 \frac{\sqrt{3}}{h}+11 \sqrt{3} h+3 \sqrt{3} h^3 & 0 & 0 & 1 & 0 & 0 \\
\end{array}
\right)\,.
\eeq}
Up to the fifth non-trivial order, $\Delta_{L=4,S=1}(h)$ and the $\bP$-functions for the $\mathfrak{sl}(2|1)$ operator with $L=4$, $K_4=2$, $K_{\bar4}=0$ are given by
{\small
\beqa
\label{eq:deltaL02}
\Delta_{L=4,S=1}(h)&=&5 + 4\, h^2  - 12\, h^4 + 68\, h^6 - 4 \left(115 + 8\, \zeta_3\right) h^8 \\
&+&\left(3332+\frac{64\, \pi ^2}{3}-\frac{136\, \pi ^4}{45}+\frac{97\, \pi ^6}{189}-\frac{127\, \pi ^8}{2160} + 416\, \zeta_3 + 320\, \zeta_5\right) h^{10} + \dots\,,\nonumber
\eeqa}
{\small
\beqa
\label{eq:PL021}
&&\bP_1(u)=\frac{1}{u^4}+\frac{4 h^2}{u^6}+\frac{2 h^4 \left(u^2+7\right)}{u^8}+h^6
   \left(\frac{48}{u^{10}}+\frac{12}{u^8}+\frac{8 \zeta_3}{u^6}-\frac{10}{u^6}\right)\\
   &&+h^8 \left(\frac{165}{u^{12}}+\frac{54}{u^{10}}+\frac{48 \zeta_3}{u^8}-\frac{60}{u^8}-\frac{80 \zeta_5}{u^6}-\frac{24 \zeta_3}{u^6}+\frac{66}{u^6}\right)+\dots\,,\nonumber\\
&&\bP_2(u)=\frac{h^2}{u^5}+\frac{5 h^4}{u^7}+\frac{20 h^6}{u^9}+h^8 \left(\frac{75}{u^{11}}+\frac{8 \zeta_3}{u^7}\right)+\dots\,,\nonumber\\
&&\bP_3(u)=-\frac{440 u^5}{189}+\frac{1208 u^3}{315}-\frac{8}{105 u^3}+\frac{92u}{45}-\frac{184}{945 u}\nonumber\\
   &&+h^2 \left(-\frac{776
   u^5}{189}-\frac{32}{105 u^5}+\frac{18692 u^3}{945}-\frac{260}{189 u^3}+\frac{748 u}{63}-\frac{1024}{945 u}\right)\nonumber\\
  &&+h^4 \left(-\frac{16}{15 u^7}+\frac{3688 u^5}{189}-\frac{148}{27 u^5}-\frac{1696 u^3}{63}-\frac{152}{35 u^3}+\frac{976 u}{135}+\frac{328}{189 u}\right)\nonumber\\ 
&&+h^6 \left(-\frac{128}{35 u^9}-\frac{896}{45 u^7}+\frac{3520 u^5 \zeta_3}{189}-\frac{64 \zeta_3}{105 u^5}-\frac{27112 u^5}{189}-\frac{17624}{945 u^5}-\frac{41984 u^3 \zeta_3}{945}\right.\nonumber\\
&&\left.-\frac{64 \zeta_3}{189 u^3}+\frac{201916 u^3}{945}+\frac{388}{945 u^3}+\frac{256 u \zeta_3}{21}+\frac{2176 \zeta_3}{189 u}+\frac{72508 u}{945}-\frac{24176}{945
   u}\right)+\dots\,,\nonumber\\
&&\bP_4(u)=-\frac{5 u^4}{3}+\frac{u^2}{3}+\frac{1}{3}+h^2 \left(-\frac{308 u^4}{27}-\frac{8}{105 u^4}+\frac{704 u^2}{21}-\frac{184}{945 u^2}-\frac{68}{45}\right)\nonumber\\
&&+h^4 \left(-\frac{8}{21 u^6}+\frac{328 u^4}{27}-\frac{274}{189 u^4}+\frac{11414 u^2}{189}-\frac{1402}{945 u^2}-\frac{1614}{35}\right)\nonumber\\
&&+h^6 \left(-\frac{32}{21 u^8}-\frac{6248}{945 u^6}-\frac{2012 u^4}{27}-\frac{5672}{945 u^4}+24 u^2 \zeta_3-\frac{8 \zeta_3}{3 u^2}\right.\nonumber\\
   &&\left.-\frac{2290 u^2}{189}-\frac{254}{189 u^2}-16 \zeta_3-\frac{27872}{945}\right)+h^8 \left(-\frac{40}{7 u^{10}}-\frac{8303}{315 u^8}-\frac{64 \zeta_3}{105 u^6}-\frac{3392}{135 u^6}\right.\nonumber\\
&&+\frac{2464 u^4 \zeta_3}{27}-\frac{1072 \zeta_3}{189 u^4}+\frac{12896
   u^4}{27}-\frac{5273}{945 u^4}-240 u^2 \zeta_5-\frac{31816 u^2 \zeta_3}{189}+\frac{80 \zeta_5}{3 u^2}\nonumber\\
   &&\left.+\frac{1000 \zeta_3}{189 u^2}+\frac{13046 u^2}{189}-\frac{14396}{945 u^2}+160
   \zeta_5-\frac{10208 \zeta_3}{63}-\frac{134012}{945}\right)+\dots\,,\nonumber
   \eeqa
   \beqa
&&\bP_5(u)=-\frac{4 u^2}{3}+\frac{1}{6 u^2}+\frac{4}{3}+h^2 \left(\frac{1}{3 u^4}-\frac{10 u^2}{7}+\frac{1}{2 u^2}+\frac{166}{15}\right)\nonumber\\
&&+h^4 \left(\frac{88}{105 u^6}+\frac{1}{u^4}+\frac{82
   u^2}{21}-\frac{2}{u^2}-\frac{264}{35}\right)\nonumber\\
&&+h^6 \left(\frac{248}{105 u^8}+\frac{241}{105 u^6}-\frac{4}{u^4}+\frac{8 \zeta_3}{3 u^2}-22 u^2+\frac{212}{21 u^2}+\frac{16 \zeta_3}{3}+\frac{6376}{105}\right)\nonumber\\
&&+h^8 \left(\frac{499}{70 u^{10}}+\frac{1177}{210 u^8}-\frac{212}{21 u^6}+\frac{16 \zeta_3}{3 u^4}+\frac{424}{21 u^4}+\frac{80 u^2 \zeta_3}{7}\right.\nonumber\\
&&\left.-\frac{80 \zeta_5}{3 u^2}-\frac{92 \zeta_3}{15 u^2}+\frac{3106 u^2}{21}-\frac{564}{7 u^2}-\frac{160 \zeta_5}{3}-\frac{2304 \zeta_3}{35}-\frac{6232}{15}\right)+\dots\,,\nonumber\\
&&\bP_6(u)=-\frac{2}{u^2}+h^2 \left(-\frac{4}{u^4}-\frac{14}{3 u^2}\right)+h^4 \left(-\frac{32}{3 u^6}-\frac{28}{3 u^4}+\frac{34}{3 u^2}\right)\nonumber\\
&&+h^6 \left(-\frac{32}{u^8}-\frac{68}{3 u^6}+\frac{68}{3
   u^4}-\frac{190}{3 u^2}\right)\nonumber\\
&&+h^8 \left(-\frac{102}{u^{10}}-\frac{62}{u^8}+\frac{56}{u^6}-\frac{380}{3 u^4}+\frac{112 \zeta_3}{3 u^2}+\frac{422}{u^2}\right)+\dots\,,\nonumber
\label{eq:PL02last}
\eeqa}
while the non-zero $c$'s used as weak coupling initial conditions read
{\small
\beqa
&&c_{1,2}=2 h^2+h^4(-10+8\zeta_3)+h^6(66-24 \zeta_3-80\zeta_5)\,,\\
&&c_{2,2}=80 h \zeta_3\,,\nonumber\\
&&c_{3,2}=\frac{604 h^3}{1575}+\frac{1282 h^5}{1575}-\frac{4484
   h^7}{945}+h^9 \left(\frac{147058}{4725}-\frac{20992 \zeta_3}{4725}\right)\,,\nonumber\\
&&+h^{11} \left(\frac{5984 \zeta_3}{675}+\frac{41984 \zeta_5}{945}-\frac{1138952}{4725}+\frac{3488 \pi ^2}{945}-\frac{26854 \pi
   ^4}{70875}+\frac{4436 \pi ^6}{59535}-\frac{19177 \pi ^8}{3402000}\right)\,,\nonumber\\
&&c_{3,4}=\frac{46 h}{225}+\frac{526 h^3}{225}+\frac{974 h^5}{225}+h^7 \left(\frac{128 \zeta_3}{105}-\frac{7102}{1575}\right)\nonumber\\
&&+h^9 \left(-\frac{9952 \zeta_3}{1575}-\frac{256 \zeta_5}{21}+\frac{6122}{315}-\frac{728 \pi
   ^2}{675}-\frac{658 \pi ^4}{10125}-\frac{4 \pi ^6}{1215}-\frac{2921 \pi ^8}{972000}\right)\,,\nonumber\\
&&c_{3,6}=-\frac{92}{4725 h}+\frac{454 h}{4725}+\frac{11866 h^3}{4725}+h^5
   \left(\frac{1088 \zeta_3}{945}+\frac{8366}{4725}\right)\nonumber\\
&&+h^7 \left(\frac{11776 \zeta_3}{4725}-\frac{2176 \zeta_5}{189}+\frac{14542}{1575}-\frac{496 \pi ^2}{2835}+\frac{12452 \pi ^4}{212625}-\frac{1618 \pi ^6}{178605}+\frac{2921 \pi ^8}{10206000}\right)\,,\nonumber\\
&&c_{3,8}=-\frac{4}{525 h^3}-\frac{62}{525 h}-\frac{44 h}{135}+h^3 \left(\frac{1186}{4725}-\frac{32 \zeta_3}{945}\right)\nonumber\\
&&+h^5 \left(\frac{9248 \zeta_3}{4725}+\frac{64 \zeta_5}{189}-\frac{11384}{4725}-\frac{16 \pi ^2}{525}+\frac{104 \pi ^4}{7875}-\frac{38 \pi ^6}{19845}+\frac{127 \pi
   ^8}{1134000}\right)\,,\nonumber\\
&&c_{3,10}=-\frac{4}{525 h^3}-\frac{244}{1575 h}+h \left(-\frac{32 \zeta_3}{525}-\frac{352}{525}\right)+h^3
   \left(-\frac{16 \zeta_3}{25}+\frac{64 \zeta_5}{105}-\frac{236}{1575}\right)\,,\nonumber\\
&&c_{3,12}=-\frac{38}{675 h}+h \left(-\frac{32 \zeta_3}{525}-\frac{2768}{4725}\right)\,,\nonumber\\
&&c_{5,2}=\frac{8}{3}+\frac{84 h^2}{5}-\frac{104 h^4}{5}+h^6 \left(\frac{32 \zeta_3}{3}+\frac{2056}{15}\right)+h^8 \left(-\frac{4608 \zeta_3}{35}-\frac{320 \zeta_5}{3}-\frac{13784}{15}\right)\,,\nonumber\\
&&c_{5,4}=\frac{1}{3 h^2}+1-\frac{20 h^2}{3}+h^4 \left(\frac{16 \zeta_3}{3}+\frac{52}{3}\right)+h^6 \left(-\frac{184 \zeta_3}{15}-\frac{160 \zeta_5}{3}-\frac{460}{3}\right)\,,\nonumber
\eeqa
\beq
c_{5,8}=\frac{1}{105 h^2}-\frac{43}{105}-\frac{4 h^2}{21}\,,\quad
c_{5,10}=-\frac{1}{3}\,,\quad
c_{6,4}=-\frac{1}{3 h^2}+\frac{1}{3}-\frac{h^2}{3}\,,\quad
c_{6,6}=-\frac{1}{3}\,. \nonumber
\eeq}

\section{Numerical results for ${\bf\Delta}$({\emph h})}
\label{app:numerics}
Our numerical results for the coefficients $c_{A, n}$ are attached to the arXiv submission as ancillary files. The  results for $\Delta(h)$ are also reported below. 
\subsection{${\bf\mathfrak{sl}}$(2)-like states}
\label{app:numericalsl2}

\begin{table}[ht]
\begin{center}
\begin{tabular}{|l|l||l|l|}
\hline
$h$ & $\Delta_{L=1,S=1}(h)$ &
$h$ & $\Delta_{L=1,S=1}(h)$
\\ \hline
0.1 & 2.077545918229727148418485943559(4) & 2.1 & 7.058259906983138858522092(6) \\  
0.2 & 2.286911738120293532704219636223(7) &  2.2 & 7.224438764422106514(1)\\ 
0.3 & 2.58411425235663174374308597422(7) & 2.3 & 7.386986434284556201(8) \\ 
0.4 & 2.931899429130996588437968238549(7) & 2.4 & 7.54613789466721195773(8) \\ 
0.5 & 3.301599812762543629406328189189(7) & 2.5 & 7.7021032437348641571(7) \\  
0.6 & 3.669001069390899384470956652261(5) & 2.6 & 7.855071188923780(1) \\
0.7 & 4.014465345366913331780856965461(5) & 2.8 & 8.152681278093727109(7) \\ 
 0.8 & 4.32787469034388191156129596828(7) & 3.0 & 8.4401529383352170(5) \\
 0.9 & 4.610485573451203638735338920432(8) & 3.2 & 8.718474731726637(2) \\ 
 1.0 & 4.869151585005447603580402087960(2) & 3.4 & 8.988482036151493(4) \\
 1.1 & 5.1103647088797283094566763014(1) & 3.6 & 9.250888767937434(7) \\ 
 1.2 & 5.338513844245525344454799468(5) & 3.8 & 9.506311011583702(6)  \\
 1.3 & 5.5563105957653673317279185(3) & 4.0 & 9.755284995227(2)   \\ 
 1.4 & 5.7654701844286937708036360(1) & 4.2 & 9.998280997074(3) \\
 1.5 & 5.9671588210612575786046727(3)  & 4.4 & 10.2357142621549(2)\\
  1.6 & 6.1622340776412210294242149(5) & 4.6 & 10.46795368429(0) \\
 1.7 & 6.351367331382348157574300(7)  & 4.8 & 10.695328791528(5)\\ 
 1.8 & 6.535107262271882908343474(7) & 5.0 & 10.918135425637(6)\\
 1.9 & 6.71391494741010817240973(4)  & 5.2 & 11.136640403794(4) \\
  2.0 & 6.88818504209752019613679(1)  & 5.6 & 11.56169005504(6)
  \\ \hline
\end{tabular}
\caption{Numerical results for $\Delta_{L=1, S=1}(h)$.}
\label{tab:DeltaL1S1}
\end{center}
\end{table}

\newpage

\begin{table}[ht]
\begin{center}
\begin{tabular}{|l|l||l|l|}
\hline
$h$ & $\Delta_{L=1,\,S=2}(h)$ &
$h$ & $\Delta_{L=1,\,S=2}(h)$
\\ \hline
 0.1 & 3.076549575409993882198041715322(8) & 1.5 & 6.301835457732233347132630(2) \\
 0.2 & 3.27358947209352367340006792421(1) & 1.6 & 6.48531812874155902633964(9) \\
 0.3 & 3.5315455139069437709269318079882(7) & 1.7 & 6.66403292579792764130221(3) \\
 0.4 & 3.8078168001033399313176033999(9)  & 1.8 & 6.83833229399729478348349233(8) \\
 0.5 & 4.0818920042378329085701415828771(1) & 1.9 & 7.00852800447317315967517(6) \\
 0.6 & 4.34637429339321283895320994206(9) & 2.0 & 7.1748972318988466455613(4) \\
 0.7 & 4.5995521010320687289391412579970(5) & 2.1 & 7.337687572423158818637(8) \\
 0.8 & 4.84172482645074795261679249873(0) & 2.2 & 7.4971211994583716275145(4) \\
 0.9 & 5.0737718609918902932825083782(7) & 2.3 & 7.6533983203215120759525(4) \\
 1.0 & 5.2966613902058939802317959277(8) & 2.4 & 7.80670006557023992676694(6) \\
 1.1 & 5.51129716729041675596673141(4) & 2.5 & 7.9571909167196777925962(3) \\
 1.2 & 5.71848010049825222569912731(1) & 2.6 & 8.10502075682286935622(5) \\
 1.3 & 5.91890762343632180844427642911(8) & 2.8 & 8.39323413454443813857(2) \\
 1.4 & 6.113184428047840495941397365(8) & 3.0 & 8.67230740916368082143(5) \\
  \hline
\end{tabular}

\caption{Numerical results for $\Delta_{L=1,\,S=2}(h)$.}
\label{tab:DeltaL1S2}
\end{center}
\end{table}

\subsection{Non-symmetric $\mathfrak{sl}$(2${\bf |}$1) states}
\label{app:numericalsl21}

\begin{table}[ht]
\begin{center}
\begin{tabular}{|l|l||l|l|}
\hline
$h$ & $\Delta_{L=2,\,S=1}(h)$ &
$h$ & $\Delta_{L=2,\,S=1}(h)$
\\ \hline
 0.1 & 3.0583138425223536525520523617144(1) & 1.5 & 6.2827823311643556042(1) \\ 
 0.2 & 3.2169917845459610847145969612802(1) & 1.6 & 6.4681645663323403799(3) \\
 0.3 & 3.4439806985878563820404952125938(9) & 1.7 & 6.6484948922621276303(1) \\ 
 0.4 & 3.709777262812282263631232340093(1) & 1.8 & 6.82418078168090376405(9) \\
 0.5 & 3.99057560301297748218692179114(6) & 1.9 & 6.99557616773107389328(6) \\ 
 0.6 & 4.26913057731143863444570240073(8) & 2.0 & 7.162991000852928720230(9) \\
 0.7 & 4.53633898038351229233235849084(6) & 2.1 & 7.326698688042131389(1) \\ 
 0.8 & 4.78986587573281131753159273216(1) & 2.3 & 7.64393759991811021781(7) \\
 0.9 & 5.03058236362588242994729784(7) & 2.5 & 7.94894521581189443819(0) \\ 
 1.0 & 5.26009487346257336212803866(4) & 2.7 &  8.2430643137204847766(8) \\
 1.1 & 5.47988231241788244483209805(7) & 2.9 & 8.5274047614439189721(8) \\
 1.2 & 5.691153863201350182789269(3) & 3.1 & 8.8028971151650518473(0) \\
 1.3 & 5.8948855908733411518817(4) & 3.3 & 9.0703311846325981633(3)  \\
1.4 & 6.0918746995284471529946(3) & 3.5 & 9.3303844544621505057(5) \\
  \hline
\end{tabular}

\caption{Numerical results for $\Delta_{L=2,\,S=1}(h)$ (with $u_4\neq u_{\bar 4}$).}
\label{tab:DeltaL2S1}
\end{center}
\end{table}

\newpage

\begin{table}[ht]
\begin{center}
\begin{tabular}{|l|l||l|l|}
\hline
$h$ & $\Delta_{L=4,\,S=1}(h)$ &
$h$ & $\Delta_{L=4,\,S=1}(h)$
\\ \hline
 0.1 & 5.0388633868644676241822546826297(8) & 1.4 & 7.2525959225896746267246(7) \\
 0.2 & 5.144185841230597954099435225377(8) & 1.5 & 7.4132015908506609153(2) \\
 0.3 & 5.293304669180183945769861210149(9) & 1.6 & 7.57064613404080615026(8) \\
 0.4 & 5.466638945852782519083338687217(1) & 1.7 & 7.72509377361324945933(7) \\
 0.5 & 5.651327339025081030520312721046(2) & 1.8 & 7.87669961712208389237(9) \\
 0.6 & 5.839840097748699283594694204296(2) & 1.9 & 8.02560874838364282201(6) \\
 0.7 & 6.02807652731490187260914627763(4) & 2.0 & 8.171956174024886529(0) \\
 0.8 & 6.21394318398974377773461287518(2) & 2.2 & 8.45745806473920852485(5) \\
 0.9 & 6.396453408629112553227909425(8) & 2.4 & 8.734102645852155605(3) \\
 1.0 & 6.57520772036382331567705855(5) & 2.6 & 9.0026632015364765359(9) \\
 1.1 & 6.7501100438531582569531237(7) & 2.8 & 9.26380917452471645(2) \\
 1.2 & 6.9212169378859605080438605(9) & 3.0 & 9.518123548274575644(1) \\
 1.3 & 7.08865832149423240961380(3) & 3.2 & 9.766117174342067304(6) \\
  \hline
\end{tabular}

\caption{Numerical results for $\Delta_{L=4,\,S=1}(h)$ (with $K_4=2,\, K_{\bar 4}=0$).}
\label{tab:DeltaL4S1}
\end{center}
\end{table}

\bibliography{Biblio3} 

\end{document}